\documentclass[12pt,a4paper]{article}
\usepackage[english]{babel}
\usepackage[numbers,comma,square,compress]{natbib}
\usepackage{amssymb,amsmath,amsfonts,bm,braket}

\usepackage{ifpdf}
\ifpdf  % Compiling PDF with pdflatex
 \usepackage[pdftex,a4paper,hmargin=28mm,vmargin=24mm]{geometry}
 \usepackage[pdftex]{hyperref}
\else  % Compiling DVI with latex
 \usepackage[dvips,a4paper,hmargin=28mm,vmargin=24mm]{geometry}
 \usepackage[dvips]{hyperref}
\fi
\hypersetup{
pdftitle={},
pdfauthor={},
}

\numberwithin{equation}{section}

\newcommand{\MP}{M_\mathrm{P}} % Planck mass
\newcommand{\R}[1][4]{{}^{(#1)}\!R}
\newcommand{\sR}{\R[3]}

\newcommand{\pb}[1]{\left\{#1\right\}}
\newcommand{\cG}{\mathcal{G}}
\newcommand{\cC}{\mathcal{C}}
\newcommand{\cH}{\mathcal{H}}

\newcommand{\cL}{\mathcal{L}}

\newcommand{\cD}{\mathcal{D}}

\newcommand{\cN}{\mathcal{N}}
\newcommand{\cB}{\mathcal{B}}
\newcommand{\cF}{\mathcal{F}}
\newcommand{\bn}{\bm{n}}

\newcommand{\nn}{\nonumber\\}

\newcommand{\email}[1]{\footnote{E-mail: \href{mailto:#1}{#1}}}

\begin{document}

\title{Canonical formulation and path integral for local vacuum energy
sequestering}

\author{R.~Bufalo$^{a}$\email{rodrigo.bufalo@ufabc.edu.br}~,
J.~Kluso\v{n}$^{b}$\email{klu@physics.muni.cz} ~ and
M.~Oksanen$^{c}$\email{markku.oksanen@helsinki.fi}
\vspace{.6em}\\
\textit{$^{a}$ \small Universidade Federal do ABC, Centro de Ci\^encias Naturais e Humanas,}\\
\textit{ \small Av. dos Estados, 5001, 09210-580 Santo Andr\'e, SP, Brazil}\\
\textit{$^{b}$ \small Department of Theoretical Physics and Astrophysics, Faculty of Science,} \\
\textit{\small Masaryk University, Kotl\'a\v{r}sk\'a 2, 611 37, Brno, Czech Republic}\\
\textit{$^{c}$ \small Department of Physics, University of Helsinki, P.O. Box 64}\\
\textit{ \small FI-00014 Helsinki, Finland}\\
}
\date{}
\maketitle

\begin{abstract}
We establish the Hamiltonian analysis and the canonical path integral
for a local formulation of vacuum energy sequestering. In particular,
by considering the state of the universe as a superposition of vacuum
states corresponding to different values of the cosmological and
gravitational constants, the path integral is extended to include
integrations over the cosmological and gravitational constants. The
result is an extension of the Ng-van Dam form of the path integral of
unimodular gravity. It is argued to imply a relation between the
fraction of the most likely values of the gravitational and cosmological
constants and the average values of the energy density and pressure of
matter over spacetime. Finally, we construct and analyze a BRST-exact
formulation of the theory, which can be considered as a topological
field theory.
\end{abstract}
\begin{flushleft}
\end{flushleft}

\section{Introduction}

According to a convincing body of observations the expansion of the
universe is accelerating. Thus, assuming that general relativity (GR)
continues to describe the universe accurately at the largest scales,
we observe that the cosmological constant is not zero. There are many
proposals seeking to explain the smallness of the cosmological constant,
although with no understanding of its fundamental origin.
Unfortunately, the situation is not that simple as one might initially
expect, because the theoretical and observational results fail to come
to a unified answer, since the theoretical appraisal exceeds the
observed value by 120 orders of magnitude. Furthermore, the cosmological
constant predicted by quantum field theory (QFT) is radiatively unstable
to the extreme. This is the so-called cosmological constant problem
\cite{Weinberg:1988cp,Padilla:2015aaa}, the worst but most important
problem of fine-tuning in physics.

The origin of this outstanding disagreement can be traced back to the
universality of gravity and the quantum generation of vacuum energy by
virtual particles. In a quantum field theorist point-of-view even the
vacuum possesses energy density, given by the resummation of the QFT
bubble diagrams. In GR, vacuum energy contributes to the cosmological
constant, and the vacuum geometry must be curved due to the equivalence
principle.

On the one hand, if one approaches this problem by enforcing a symmetry
principle, for instance supersymmetry and/or conformal symmetry, the
huge vacuum energy could be canceled. However, at scales below a TeV,
these symmetries are broken. On the other hand, one could alternatively
approach this situation by means of a dynamical adjustment of vacuum
energy, where a nongravitating degree of freedom is responsible for
``eating'' it all up. The major issue with this idea is to work around
Weinberg's no-go theorem \cite{Weinberg:1988cp} which prohibits such
adjustment in any standard QFT coupled to gravity.

One of the first and best-known (minimal) modifications of GR that was
hoped to shed new light on the cosmological constant problem is
unimodular gravity \cite{Einstein:1919gv,vanderBij:1981ym,
Buchmuller:1988wx,Padilla:2014yea}.
It is well known that the field equation for the metric in unimodular
gravity is either the traceless Einstein equation or, thanks to the
Bianchi identity, the Einstein equation with a cosmological constant
\cite{Unruh:1988in}. Actually, the main conceptual difference to GR is
that the cosmological constant of unimodular gravity is a constant of
integration, rather than a coupling constant.
This different point of view on the cosmological constant has led to
considerable interest in several topics within the unimodular gravity
scenario; see Refs.~\cite{Henneaux:1989zc,Ng:1990rw,Ng:1990xz,Kuchar:1991xd,
Smolin:2009ti,Bufalo:2015wda} and references therein. Unfortunately, a
similar problem with the renormalization or fine-tuning of the
cosmological constant is found as in GR
\cite{Weinberg:1988cp,Padilla:2015aaa}.

Among the several new proposals made in recent years, which attempt to
resolve the cosmological constant problem via (minimal) modification of
GR, we would like to call attention to a particular proposal called
vacuum energy sequestering \cite{Kaloper:2013zca}. It includes
a global mechanism for decoupling the vacuum energy generated by matter
loops from gravity. Hence it appears to be able to evade Weinberg's
no-go theorem.

In order to avoid the drawbacks of previous approaches -- mainly the
extreme sensitivity of the (diffeomorphism-allowed) contribution to
the cosmological constant for any change of the matter sector parameters
or addition of higher-order loop corrections, and the need for it to be
tuned by hand order by order in perturbation theory to ensure a
particular finite value of the cosmological constant -- the vacuum energy
sequestering theory provides \emph{by design} that all quantum-generated
vacuum energy contributions from a protected matter sector cancel
completely from the gravitational equations of motion
\cite{Kaloper:2013zca,Kaloper:2014dqa,Kaloper:2014fca}. Thus, the only
vacuum energy remaining that now sources gravity is a renormalized
vacuum energy, which is automatically radiatively stable and fully
consistent with the concept of renormalization in QFT. The sequestering
mechanism works at each and every order in perturbation theory, so there
is no need to retune the classical cosmological constant when
higher-order loop corrections are included. Hence, one is left with a
radiatively stable cosmological constant, which is completely
independent of the vacuum energy contributions from the protected matter
sector \cite{Kaloper:2014dqa}.

The core of this proposal may be thought of as understanding GR as a hybrid
of unimodular gravity in conjunction with variational procedures
\cite{Lanczos} to fix values of global variables such as the
cosmological constant.
The main feature of the sequestering mechanism is engendered by means of
a global term $\sigma\left( \frac{\lambda}{\eta^4\mu^4} \right)$,
which is added to the Einstein-Hilbert action [see Eq.~\eqref{SVESglobal}],
so that all scales in the matter sector are now functionals of the gauge-invariant four-volume
element of the universe $\int d^4x\sqrt{-g}$
\cite{Kaloper:2013zca}.
The prescription of the vacuum energy sequestering mechanism uses a
global scaling symmetry, in terms of a scaling parameter measuring the
matter sector scales in Planck units, as an organizing principle for
accounting for all quantum vacuum energy contributions.
This is the key point of the sequestering mechanism that provides the
way around Weinberg's no-go theorem. Unlike in GR or in its
unimodular formulation, now the four-volume $\int d^4x\sqrt{-g}$ is an
independent variable. Remarkably, locally the sequestering theory
behaves just like standard GR, in the (semi) classical limit, but
without a large cosmological constant and without its radiative
instability \cite{Kaloper:2014dqa}.

Although successfully explaining the net value and stability of the
cosmological constant through a dynamical adjustment, the
global term in the sequestering mechanism is unusual
and appears to conflict with the expectations about the microscopic
origin of the mechanism.
In order to address this deficiency, a local formulation of the theory
has been recently proposed \cite{Kaloper:2015jra,Kaloper:2016yfa}.
In this manifestly local version of the sequestering mechanism
the global terms are regarded as conserved quantities,
and gauge redundancies are introduced in order to allow for the rigid
variables to become solutions of local field equations.
Actually, this local setup might be obtained from the global one by
using a similar reparametrization invariance approach as in the
Henneaux-Teitelboim form of unimodular gravity
\cite{Henneaux:1989zc}.

Interestingly, by virtue of the local conservation laws, the
modifications of the gravitational sector in the local sequestering
mechanism behave very similarly to the global setup of
\cite{Kaloper:2013zca}.
However, now solutions display the new features of a finite, eternal
cosmological constant, and the spacetime
volume of the underlying geometry does not have to be finite, while
supporting a finite Planck scale and (protected) matter sector scales.
Recently, the local framework has been further explored
to understand cosmological behavior, the effects of phase
transitions, and the interplay between gravity and particle
physics \cite{Kaloper:2016yfa}.

It has been speculated that the local setup of the sequestering
mechanism admits a standard Hamiltonian dynamics, allowing thus a
definition of the Feynman path integral. We therefore wish to establish
a Hamiltonian formalism and determine the canonical path
integral for the local version of vacuum energy sequestering.
The path integral enables us to work out an interesting point regarding
the gravitational and cosmological constants. The situation is similar
to the case of unimodular gravity, where an additional integration over
the cosmological constant can be included, meaning that we are
integrating over different physical boundary conditions (vacua)
\cite{Ng:1990rw,Ng:1990xz,Bufalo:2015wda}.
We now extend the idea of the Ng--van Dam form of the path integral
to encompass integration over both the cosmological and gravitational
constants in the path integral of the local sequestering model.

We also explore the possibility of a topological or Becchi-Rouet-Stora-Tyutin (BRST)-exact
formulation of the local vacuum energy sequestering model. This
approach is inspired by a recent work on a so-called topological induced
gravity \cite{Oda:2016xls}, which is shown to be a simple special case
of the local formulation of vacuum energy sequestering in
Sect.~\ref{secIG}, and by a similar approach to unimodular gravity in
\cite{Nojiri:2016mlb} (see Sect.~\ref{secNo}). We regard that the
gravitational action of vacuum energy sequestering appears as a gauge-fixing action along with an appropriate ghost action. As a result the
action of vacuum energy sequestering becomes BRST exact and can be
considered as a topological field theory \cite{Witten:1988ze}. The
canonical structure of the ghost sector and the BRST charges are fully
analyzed.

The paper is organized as follows.
In Sect.~\ref{secIntVES} we present the vacuum energy sequestering
theory in the original (global) form and its local extension, and
explain how the sequestering mechanism works.
Section \ref{secCanVES} is dedicated to the Hamiltonian formulation
of the local theory.
In Sect.~\ref{secPI} the canonical path integral is established and we
show that it can written in an extended Ng-van Dam form. That
enables us to obtain a relation between the cosmological and
gravitational constants.
In Sect.~\ref{secIG} we prove a relation between the topological induced
gravity as a particular case of the local vacuum energy sequestering
model.
Section \ref{secIG-BRST} conceives the local sequestering model in a
BRST-exact form, and a canonical analysis of the resulting theory is
performed.
For the purpose of comparison, we discuss another attempt to cope with
the cosmological constant problem \cite{Nojiri:2016mlb} in
Sect.~\ref{secNo}, which includes the cosmological constant in terms of
a topological field theory, but lacks a mechanism for ensuring its
radiative stability.
Final remarks are presented in Sect.~\ref{conc}.

\section{Vacuum energy sequestering}
\label{secIntVES}

\subsection{Global mechanism}
\label{secIntVESglobal}

The vacuum energy sequestering mechanism is based on the presence of two
rigid variables with no local degrees of freedom: the bare cosmological
constant $\lambda$ and a scaling parameter $\eta$ measuring the matter
sector scales in Planck units.
The main result of this procedure is that it sets the boundary condition
upon $\eta$ in such a way that at every order of the loop expansion it
takes precisely the necessary value in order to completely cancel the
vacuum energy contribution from the matter sector at that order.
The action for the given theory can be expressed as
\cite{Kaloper:2013zca}
\begin{equation}\label{SVESglobal}
 S=\int d^4x\sqrt{-g}\left( \frac{\MP^2}{2}R -\lambda
 +\eta^4 \cL_\mathrm{m}(\eta^{-2} g_{\mu\nu},\Psi) \right)
 +\sigma\left( \frac{\lambda}{\eta^4\mu^4} \right),
\end{equation}
where the global function $\sigma$ is required to be an odd
differentiable function, and the mass scale $\mu$ is around the QFT
cutoff. Matter fields -- denoted collectively by $\Psi$ -- are
coupled minimally to the metric.

From the variation of Eq.~\eqref{SVESglobal} with respect to $\lambda$,
which links the four-volume to the scaling parameter $\eta$, we find  a
necessary condition for the matter scales to be nonzero, representing
the picture of a finite universe in spacetime, collapsing in the future
\cite{Kaloper:2014fca}.
In particular, if the matter sector is the Standard Model of elementary
particles, the vacuum energy sequestering mechanism prevents it from
generating large contributions to the net cosmological constant.
The variation of Eq.~\eqref{SVESglobal} upon the metric $g_{\mu\nu}$ yields
\begin{equation}\label{eomVES}
\MP^2 G^\mu _{~\nu}= T^\mu _{~\nu}-\frac{1}{4}\delta^\mu
_{~\nu}\left\langle T^\alpha _{~\alpha} \right\rangle ,
\end{equation}
where $T^\mu _{~\nu}$ is the energy-momentum tensor of the matter
fields, and we have eliminated $\lambda$ by its constraint equation
$\lambda= \frac{1}{4}\left\langle T^\alpha _{~\alpha} \right\rangle$,
where we have defined the four-volume average by $\left\langle P
\right\rangle =\int d^4x\sqrt{-g}P /\int d^4x\sqrt{-g} $.
Now we see that Eq.~\eqref{eomVES} is the key result of the sequestering
mechanism. Furthermore, this field equation is unlike in
unimodular gravity, where the restricted variation removes the trace
equation that involves the vacuum energy, but returns it as an arbitrary
integration constant.
On the other hand, in the sequestering mechanism there are no hidden
equations or integration constants, all the sources are automatically
accounted for in the right-hand side of Eq.~\eqref{eomVES}.

Let us now scrutinize the main feature from \eqref{eomVES}.
Remarkably, the matter-sector quantum corrections to vacuum energy are
all accounted for in the average $\left\langle T^\alpha _{~\alpha}
\right\rangle$, and cancel precisely from the right-hand side of Eq.~
\eqref{eomVES}.
This can be better viewed by extracting the constant contribution into
the stress energy, $V_{\mathrm{vac}}$, so that we can rewrite the stress
energy tensor as $T^\mu _{~\nu}= -V_{\mathrm{vac}}\delta^\mu
_{~\nu}+\tau^\mu _{~\nu}$, where the tensor $\tau^\mu _{~\nu}$ describes
local excitations. We therefore see that by means of such decomposition
$V_{\mathrm{vac}}$ completely drops out from \eqref{eomVES}
\cite{Kaloper:2013zca,Kaloper:2014dqa}.
Furthermore, this result shows that the only vacuum energy that sources
gravity is a renormalized vacuum energy, which is automatically
radiatively stable.

However, note that there is a residual cosmological constant left: the
historic average $\left\langle \tau^\alpha _{~\alpha} \right\rangle$
which is notably insensitive to vacuum loop corrections, and is
precisely small in large and old universes by virtue of two approximate
symmetries, the scalings $\eta \rightarrow \Omega \eta$,
$g_{\mu\nu} \rightarrow \Omega^{-2}g_{\mu\nu}$, and $\lambda \rightarrow
\Omega^4 \lambda$, and the shifts	 $\lambda \rightarrow \lambda+\alpha
\eta^4$ and $\cL_\mathrm{m} \rightarrow \cL_\mathrm{m}-\alpha$, which
are broken only by the gravitational sector \cite{Kaloper:2014dqa}. At
last, as with any leftover of a UV-sensitive physical quantity in QFT, the
numerical value of the finite part of the cosmological constant is not
determined by the theory, but rather determined to match observations.

\subsection{Local formulation}
\label{secIntVESlocal}

A local formulation of the sequestering mechanism has been proposed
\cite{Kaloper:2015jra} mainly to deal with the microscopic origin of the
mechanism, interpreting the
global terms $\sigma\left( \frac{\lambda}{\eta^4\mu^4} \right)$ as
conserved quantities,
so that the rigid variables $\lambda$, $\eta$ are solutions of local
field equations.
As a first step on its definition, it is relevant for
definiteness of the local formulation to absorb $\eta$ into the
definition of the Planck scale in the action \eqref{SVESglobal}. This
is achieved by the change of variables $g_{\mu\nu} \rightarrow
\frac{\kappa^2}{\MP^2}g_{\mu\nu} $, $\lambda\rightarrow \left(\frac{\MP
^2}{\kappa^2}\right)^2\lambda $
 where a new variable $\kappa^2=\frac{\MP	^2}{\eta^2}$ is defined.
In terms of this new variable, the action \eqref{eomVES} now reads
\begin{equation}\label{SVESglobal2}
 S=\int d^4x\sqrt{-g}\left( \frac{\kappa^2}{2}R -\lambda
 +  \cL_\mathrm{m}( g_{\mu\nu},\Psi) \right)
 +\sigma\left( \frac{\lambda}{\mu^4} \right).
\end{equation}
The variation of $\eta$ is now replaced by the $\kappa^2$ parameter, but
the sequestering mechanism obtained from the field equations remains
intact \cite{Kaloper:2015jra}.

Now the path chosen to promote the rigid parameters $\kappa^2$,
$\lambda$ to local fields, and reinterpret the
global term as an integral of local expressions, which simultaneously
yield local equations
$\partial_\mu\kappa^2=0$, $\partial_\mu\lambda=0$
is very similar to the known gauge-invariant
formulation of unimodular gravity by Henneaux and Teitelboim
\cite{Henneaux:1989zc}.
In their formulation, the unimodular constraint
$\sqrt{-g}=\varepsilon_0$ is replaced by a diffeomorphism-invariant form $\sqrt{-g} =\partial_\mu\tau^\mu$ enforced
by a term $-\int d^4 \lambda(x) \left( \sqrt{-g} -\partial_\mu\tau^\mu
\right)$,
where $\lambda(x)$ is treated as a Lagrange multiplier satisfying
$\partial_\mu\lambda=0$ and $\tau^\mu$ is a vector density.
This diffeomorphism-invariant form can be derived from the global one
via a parametrization of space-time coordinates \cite{Kuchar:1991xd} (see
Refs.~\cite{Lanczos,Dirac} for a review).

Now, following a similar path for the local sequestering model, we wish
to replace the $\lambda$-dependent terms in Eq.~\eqref{SVESglobal2} with a (diffeomorphism-preserving) gauge-fixing
action
\begin{equation}\label{SVESgf}
 S_{gf}=-\int \left(\lambda \sqrt{-g}d^4x -
d\hat{A}\hat{\sigma}\left(\frac{\lambda}{\mu^4} \right) \right),
\end{equation}
where  $d\hat{A}$ is the exterior derivative of an auxiliary three-form
$\hat{A}$ (see below).
The new local additions should not gravitate directly in order to
preserve the main feature of sequestering.
On the one hand, the field equation obtained by varying $\hat{A}$ is
precisely $\partial_\mu \lambda=0$, fixing the Lagrange multiplier
$\lambda(x)$ to be an arbitrary rigid contribution to the total
cosmological constant.
On the other hand, the variation with
respect to $\lambda(x)$ yields that $\hat{A}$ is a nonpropagating,
auxiliary field.
It should be clear that the real reason for the absence of any local
degrees of freedom from $\lambda(x)$ is the newly introduced gauge
redundancy of the three-form (see discussion below).

Since we have one more rigid Lagrange multiplier in Eq.~\eqref{SVESglobal2},
$\kappa^2$, we can follow exactly the same procedure to make it local
off shell, and constant on shell by means of an extra copy of
Eq.~\eqref{SVESgf}.
So the action for local vacuum energy sequestering
\cite{Kaloper:2015jra} can be written as (cf. Ref.~\cite{Kaloper:2015jra})
\begin{equation}\label{SVES}
\begin{split}
 S&=\int d^4x\sqrt{-g}\left( \frac{\kappa^2}{2}R -\lambda
 +\cL_\mathrm{m}(g_{\mu\nu},\Psi) \right)
 +\int dA \sigma\left( \frac{\kappa^2}{\MP^2} \right)
 +\int d\hat{A} \hat\sigma\left( \frac{\lambda}{\mu^4} \right),
\end{split}
\end{equation}
where the bare gravitational and cosmological constants $(\kappa^2$ and
$\lambda$) are local fields.  $dA$ and $d\hat{A}$ are exterior
derivatives of two auxiliary three-forms $A$ and $\hat{A}$. Furthermore,
$\sigma$ and $\hat\sigma$ are two smooth functions, where $\MP$
and $\mu$ are the (cutoff) energy scales associated with gravity and
matter, respectively.  The measure four-forms can be
written as
\begin{align}
 dA&=d^4x (*dA)
 =d^4x \frac{1}{4!}\epsilon^{\mu\nu\rho\sigma}dA_{\mu\nu\rho\sigma}
 =d^4x \partial_\mu \omega^\mu,\\
 d\hat{A}&=d^4x (*d\hat{A})
 =d^4x \frac{1}{4!}\epsilon^{\mu\nu\rho\sigma}
 d\hat{A}_{\mu\nu\rho\sigma}
 =d^4x \partial_\mu \tau^\mu,
\end{align}
where $\omega^\mu$ and $\tau^\mu$ are vector densities of unit weight:
\begin{equation}\label{omegatau}
 \omega^\mu=\frac{1}{3!}\epsilon^{\mu\nu\rho\sigma}A_{\nu\rho\sigma},
 \quad \tau^\mu=\frac{1}{3!}\epsilon^{\mu\nu\rho\sigma}
 \hat{A}_{\nu\rho\sigma}.
\end{equation}
The four-forms are invariant under a gauge transformation $A\rightarrow
A+dB$ and $\hat{A}\rightarrow \hat{A}+d\hat{B}$, where $B$ and
$\hat{B}$ are arbitrary two-forms. Equivalently, the gauge
transformations of the measure can be written for the vector densities
\eqref{omegatau} as
$\omega^\mu\rightarrow\omega^\mu+\partial_\nu\alpha^{\mu\nu}$ and
$\tau^\mu\rightarrow\tau^\mu+\partial_\nu\hat\alpha^{\mu\nu}$, where
$\alpha^{\mu\nu}$ and $\hat\alpha^{\mu\nu}$ are antisymmetric tensor
densities of unit weight. In the presence of a metric we can further
write $\omega^\mu=\sqrt{-g}U^\mu$ and $\tau^\mu=\sqrt{-g}V^\mu$, so that
(similar to the situation in the fully diffeomorphism-invariant formulation
of unimodular gravity\cite{Bufalo:2015wda})
\begin{equation}
 \partial_\mu \omega^\mu=\sqrt{-g}\nabla_\mu U^\mu,\quad
 \partial_\mu \tau^\mu=\sqrt{-g}\nabla_\mu V^\mu,
\end{equation}
where $U^\mu$ and $V^\mu$ are vector fields.
Hence we rewrite the action for local vacuum energy sequestering
\eqref{SVES} as
\begin{equation}\label{SVES2}
 S=\int d^4x\sqrt{-g}\biggl[ \frac{\kappa^2}{2} R -\lambda +\sigma\left(
 \frac{\kappa^2}{\MP^2}  \right) \nabla_\mu U^\mu
 +\hat\sigma\left( \frac{\lambda}{\mu^4} \right) \nabla_\mu V^\mu
 +\cL_\mathrm{m}(g_{\mu\nu},\Psi) \biggr].
\end{equation}

\section{Canonical formulation of the local theory of vacuum energy
sequestering}
\label{secCanVES}

\subsection{ADM representation of the action}
The canonical formulation of the original vacuum energy sequestering model
\eqref{SVESglobal} \cite{Kaloper:2013zca} (see also Ref.~\cite{Kaloper:2014dqa}),
where the sequestering of vacuum energy is achieved by including a
(scaling) function outside of the action,
has already been considered in Ref.~\cite{Kluson:2014tma}.
For a canonical formulation of the local version of vacuum energy
sequestering it is convenient to further rewrite the action
\eqref{SVES2} by partial integration as
\begin{equation}\label{SVES3}
 S=\int d^4x\sqrt{-g}\biggl[ \frac{\kappa^2}{2} R -\lambda
 - \sigma'\left( \frac{\kappa^2}{\MP^2} \right)
 \frac{U^\mu\nabla_\mu\kappa^2}{\MP^2}
 -\hat\sigma'\left( \frac{\lambda}{\mu^4} \right)
 \frac{V^\mu\nabla_\mu\lambda}{\mu^4}
 +\cL_\mathrm{m}(g_{\mu\nu},\Psi) \biggr],
\end{equation}
where $\sigma'$ and $\hat\sigma'$ denote the first derivatives of the
smooth functions $\sigma$ and $\hat\sigma$, which are assumed to be
nonvanishing. In the simplest permitted case $\sigma'$ and $\hat\sigma'$
are constant (see Sec.~\ref{secIG}). Since we are mostly
interested in the propagating degrees
of freedom and the role of the gravitational and cosmological constants, we
omit all boundary terms. However, for the given form of the action
\eqref{SVES3} the boundary terms are of the same form as in GR (as well
as in unimodular gravity \cite{Bufalo:2015wda}).

An Arnowitt-Deser-Misner (ADM) representation of the gravitational part
of the action \eqref{SVES3} is obtained as
\begin{multline}\label{S.ADM}
 S_\mathrm{ADM}[N,N^i,h_{ij},\kappa^2,U_{\bn},U^i,\lambda,V_{\bn},V^i]
 =\int dt d^3x N\sqrt{h}\bigg[ \frac{\kappa^2}{2}
 \left( K_{ij}\cG^{ijkl}K_{kl}+\sR \right)\\
 -\left( K-\frac{\sigma'}{\MP^2}U_{\bn} \right)\nabla_n\kappa^2
-D_iD^i\kappa^2 -\frac{\sigma'}{\MP^2}U^i\partial_i\kappa^2\\
 -\lambda +\frac{\hat\sigma'}{\mu^4}V_{\bn}\nabla_n\lambda
 -\frac{\hat\sigma'}{\mu^4}V^i\partial_i\lambda \bigg],
\end{multline}
where the arguments of the functions $\sigma'$ and $\hat\sigma'$ are
omitted, but their dependence on $\kappa^2$ and $\lambda$,
respectively, should be kept in mind. The vector fields have been
decomposed to components normal and tangent to the spatial
hypersurface $\Sigma_t$, defined as
\begin{equation}
 U_{\bn}=n_\mu U^\mu,\quad  U^i=(\delta^i_{\ \mu}+n^i n_\mu)U^\mu,\quad
 n^\mu=\left( \frac{1}{N},\frac{N^i}{N} \right),
\end{equation}
where $n^\mu$ is the unit normal to $\Sigma_t$.\footnote{Tensors and
tensor densities that are tangent to the spatial hypersurfaces are
denoted with latin indices $(i.j.\ldots)$ which run from 1 to 3. For a
more detailed description of the notation see Ref.~\cite{Bufalo:2015wda}.}
The canonical formulation of the matter action is identical to that of GR.

\subsection{Hamiltonian and constraints}
Canonical momenta conjugate to $N$, $N^i$, $h_{ij}$, $\kappa^2$,
$U_{\bn}$, $U^i$, $\lambda$, $V_{\bn}$ and $V^i$ are denoted by $\pi_N$,
$\pi_i$, $\pi^{ij}$, $P_{\kappa^2}$, $P_{\bn}$, $P_i$, $p_\lambda$,
$p_{\bn}$ and $p_i$, respectively. We obtain the primary constraints
\begin{equation}
 \pi_N\approx0,\quad \pi_i\approx0,\quad P_{\bn}\approx0,\quad
 P_i\approx0,\quad p_{\bn}\approx0,\quad p_i\approx0.
\end{equation}
and
\begin{equation}\label{cClambda}
 \cC_\lambda=p_\lambda
 -\sqrt{h}\frac{\hat\sigma'}{\mu^4}V_{\bn}\approx0.
\end{equation}
The momenta conjugate to the metric $h_{ij}$ and $\kappa^2$ are defined
as
\begin{equation}\label{pi^ij}
 \pi^{ij} =\frac{1}{2}\sqrt{h}\kappa^2\cG^{ijkl}K_{kl}
 -\frac{1}{2}\sqrt{h}h^{ij}\nabla_n\kappa^2
\end{equation}
and
\begin{equation}\label{pi_kappa}
 P_{\kappa^2}=-\sqrt{h}\left( K-\frac{\sigma'}{\MP^2}U_{\bn} \right).
\end{equation}
The time derivatives of $\kappa^2$ and $h_{ij}$ are solved from
Eqs.~\eqref{pi^ij} and \eqref{pi_kappa} as
\begin{equation}
 \nabla_n\kappa^2=\frac{1}{N}\left( \partial_t\kappa^2
 -N^i\partial_i\kappa^2  \right)
 =-\frac{2}{3\sqrt{h}}\left( \pi -\kappa^2 P_{\kappa^2}
 +\sqrt{h}\frac{\kappa^2}{\MP^2}\sigma' U_{\bn} \right)
\end{equation}
and
\begin{equation}
 K_{ij}=\frac{1}{2N}\left( \partial_th_{ij}-2D_{(i}N_{j)} \right)
 =\frac{2\cG_{ijkl}\pi^{kl}}{\sqrt{h}\kappa^2}
 +\frac{h_{ij}}{3\sqrt{h}\kappa^2}\left( \pi -\kappa^2 P_{\kappa^2}
 +\sqrt{h}\frac{\kappa^2}{\MP^2}\sigma' U_{\bn} \right),
\end{equation}
where $\pi=h_{ij}\pi^{ij}$.

The Hamiltonian is obtained as
\begin{equation}\label{H}
 H=\int d^3x\left( N\cH_T+N^i\cH_i +u_N\pi_N+u_N^i\pi_i
 +u_\lambda \cC_\lambda +u_{\bn}P_{\bn}+u^iP_i
 +v_{\bn}p_{\bn} +v^ip_i \right),
\end{equation}
where the so-called super-Hamiltonian and supermomentum are
defined as
\begin{equation}\label{cHT}
\begin{split}
 \cH_T&=\frac{2\pi^{ij}\cG_{ijkl}\pi^{kl}}{\sqrt{h}\kappa^2}
 +\frac{1}{3\sqrt{h}\kappa^2}\left( \pi -\kappa^2 P_{\kappa^2}
 +\sqrt{h}\frac{\kappa^2}{\MP^2}\sigma' U_{\bn} \right)^2
 -\frac{1}{2}\sqrt{h}\kappa^2\sR \\
 &\quad +\sqrt{h} \left( D_iD^i\kappa^2
 +\frac{\sigma'}{\MP^2}U^i\partial_i\kappa^2 +\lambda
+\frac{\hat\sigma'}{\mu^4} V^i\partial_i\lambda \right)
\end{split}
\end{equation}
and
\begin{equation}\label{cHi}
 \cH_i=-2h_{ij}D_k\pi^{jk} +\partial_i\kappa^2 P_{\kappa^2}
 +\partial_i\lambda p_\lambda,
\end{equation}
respectively, where we introduced the inverse De Witt metric as
\begin{equation}
 \cG_{ijkl}=\frac{1}{2}(h_{ik}h_{jl}+h_{il}h_{jk})
 -\frac{1}{2}h_{ij}h_{kl},
\end{equation}
and $u_N,u_N^i,u_\lambda,u^i,u_{\bn},v^i,v_{\bn}$ are unspecified
Lagrange multipliers for the primary constraints.

Each primary constraint must be preserved under time evolution.
For $\pi_N\approx0$ and $\pi_i\approx0$ we obtain the
Hamiltonian constraint
\begin{equation}
 \cH_T\approx0
\end{equation}
and the momentum constraint
\begin{equation}
 \cH_i\approx0.
\end{equation}
We extend the momentum constraint \eqref{cHi} with terms that are
proportional to the primary constraints $P_{\bn}$ and $p_{\bn}$ so that
the momentum constraint generates spatial diffeomorphisms for
all the variables that are involved in the constraints.\footnote{We do
not include a generator for the variables $\left(U^{i},P_{j}\right)$
and $\left(V^{i},p_{j}\right)$ since the terms of the Hamiltonian
constraints \eqref{cHT} that depend on $U^{i}$ and $V^{i}$
are proportional to the constraints \eqref{cBi} and \eqref{cCi},
respectively.}
For that reason we redefine
\begin{equation}
 \cH_i=-2h_{ij}D_k\pi^{jk} +\partial_i\kappa^2 P_{\kappa^2}
 +\partial_i\lambda p_\lambda +\partial_iU_{\bn}P_{\bn}
 +\partial_iV_{\bn}p_{\bn}\approx0.
\end{equation}
It is useful to define global (smeared) versions of the Hamiltonian and
momentum constraints:
\begin{equation}\label{HMC}
 \cH_T[\xi]=\int d^3x\xi\cH_T,\quad
 \Phi[\chi^i]=\int d^3x \chi^i\cH_i.
\end{equation}
The preservation of the constraints $P_i\approx0$ and $p_i\approx0$,
is ensured by introducing the secondary constraints
\begin{align}
 \cB_i&=\partial_i\kappa^2\approx0,\label{cBi}\\
 \cC_i&=\partial_i\lambda\approx0.\label{cCi}
\end{align}
These constraints imply that $\kappa^2$ and $\lambda$ are constant
across space. We define smeared forms of $\cB_i$ and $\cC_i$ as
\begin{equation}\label{cBcC}
 \cB[\chi^i]=\int d^3x\chi^i\partial_i\kappa^2,\quad
 \cC[\chi^i]=\int d^3x\chi^i\partial_i\lambda.
\end{equation}
These constraints are included in the Hamiltonian \eqref{H} with
Lagrange multipliers as $\cB[u_\kappa^i]$ and $\cC[v_\lambda^i]$;
furthermore, the terms of $\cH_T$ that are proportional to $\cB_i$ and
$\cC_i$ are absorbed into the constraints $\cB[u_\kappa^i]$ and
$\cC[v_\lambda^i]$ of the Hamiltonian.
The preservation of the constraint $P_{\bn}\approx0$,
\begin{equation}
 \partial_tP_{\bn}=\pb{P_{\bn},H}\approx-\frac{2N\sigma'}{3\MP^2}
 \left( \pi -\kappa^2 P_{\kappa^2} +\sqrt{h}\frac{\kappa^2}{\MP^2}
 \sigma' U_{\bn} \right)  \approx0,
\end{equation}
requires a new secondary constraint,
\begin{equation}\label{Pi}
 \Pi=\pi-\kappa^2 P_{\kappa^2} +\sqrt{h}\frac{\kappa^2}{\MP^2}
 \sigma' U_{\bn} \approx0.
\end{equation}
The preservation of the constraint $p_{\bn}\approx0$,
\begin{equation}
 \partial_tp_{\bn}=\pb{p_{\bn},H}\approx\sqrt{h}
 \frac{\hat\sigma'}{\mu^4}u_\lambda\approx0,
\end{equation}
is ensured by fixing the Lagrange multiplier $u_\lambda$ of the
constraint $\cC_\lambda$ as
\begin{equation}\label{vlambda}
 u_\lambda=0.
\end{equation}
The preservation of the constraint \eqref{cClambda},
\begin{equation}
 \partial_t\cC_\lambda=\pb{\cC_\lambda,H}\approx
 \pb{\cC_\lambda,\cH_T[N]}-\sqrt{h}\frac{\hat\sigma'}{\mu^4}v_{\bn}
 +\pb{\cC_\lambda,\cC[v_\lambda^i]} \approx0,
\end{equation}
is ensured by fixing the Lagrange multiplier $v_{\bn}$ of the
constraint $p_{\bn}$ as
\begin{equation}
 v_{\bn}=-N\frac{\mu^4}{\hat\sigma'}
 +\frac{N\pi V_{\bn}}{\sqrt{h}\kappa^2}
 +\frac{\partial_iv_\lambda^i}{\sqrt{h}}\frac{\mu^4}{\hat\sigma'}.
\end{equation}
At this point the Hamiltonian is written as
\begin{multline}\label{H2}
 H=\int d^3x\left( N\cH_T'+N^i\cH_i +u_N\pi_N+u_N^i\pi_i
 +u_{\bn}P_{\bn}+u^iP_i +v^ip_i +u_\Pi\Pi \right.\\
 +\left.u_\kappa^i\cB_i +v_\lambda^i\cC_i' \right),
\end{multline}
where we have defined the constraints
\begin{align}
 \cH_T'&=\cH_T -\frac{\mu^4}{\hat\sigma'}p_{\bn}
 +\frac{\pi V_{\bn}}{\sqrt{h}\kappa^2}p_{\bn}
 \approx0,\label{cHT'}\\
 \cH_T&=\frac{2\pi^{ij}\cG_{ijkl}\pi^{kl}}{\sqrt{h}\kappa^2}
 -\frac{1}{2}\sqrt{h}\kappa^2\sR + \sqrt{h}D_iD^i\kappa^2
 +\sqrt{h}\lambda \approx0,\label{cHT2}\\
 \cC_i'&=\cC_i -\mu^4\partial_i\left(
 \frac{p_{\bn}}{\sqrt{h}\hat\sigma'} \right)
 =\partial_i\lambda -\mu^4\partial_i\left(
 \frac{p_{\bn}}{\sqrt{h}\hat\sigma'} \right)
 \approx0, \label{cCi'}
\end{align}
and $u_N,u_N^i,u_{\bn},u^i,u_\Pi,u_\kappa^i,v^i,v_\lambda^i$ are
unspecified Lagrange multipliers. Note that we have included the
constraint \eqref{Pi} with a Lagrange multiplier as $\Pi[u_\Pi]=\int
d^3x u_\Pi\Pi$ and absorbed the terms proportional to $\Pi$ from the
Hamiltonian constraint. Rewriting the consistency condition for
$P_{\bn}$ with the Hamiltonian \eqref{H2},
\begin{equation}
 \partial_tP_{\bn}\approx\pb{P_{\bn},\Pi[u_\Pi]}
 =-\sqrt{h}\frac{\kappa^2}{\MP^2}\sigma' u_\Pi=0,
\end{equation}
implies
\begin{equation}
 u_\Pi=0,
\end{equation}
i.e., the constraint $\Pi$ drops out of the Hamiltonian.

Then we must establish the preservation of the secondary constraints
$\cH_T$, $\cH_i$, $\cB_i$, $\cC_i$ and $\Pi$. First we consider
preservation of $\Pi$,
\begin{equation}
 \partial_t\Pi=\pb{\Pi,H}\approx\pb{\Pi,\cH_T[N]}
 +\sqrt{h}\frac{\kappa^2}{\MP^2}\sigma' u_{\bn}
 +\pb{\Pi,\cB[u_\kappa^i]} \approx0,
\end{equation}
which is achieved by fixing the Lagrange multiplier $u_{\bn}$ of the
constraint $P_{\bn}$ as
\begin{equation}
 u_{\bn}=N\frac{\MP^2}{\kappa^2 \sigma'}\left( \frac{3}{2}D_iD^i\kappa^2
 +2\lambda  \right) +N\frac{\pi U_{\bn}}{\sqrt{h}\kappa^2}
 +\frac{3\MP^2}{2\kappa^2 \sigma'} h^{ij}\partial_iN\partial_j\kappa^2
 +\frac{\MP^2}{\sqrt{h}\sigma'}\partial_iu_\kappa^i.
\end{equation}
The Hamiltonian is written as
\begin{equation}\label{H3}
 H=\int d^3x\left( N\cH_T''+N^i\cH_i +u_N\pi_N+u_N^i\pi_i
 +u^iP_i +v^ip_i +u_\kappa^i\cB_i' +v_\lambda^i\cC_i' \right),
\end{equation}
where we have defined a first-class Hamiltonian constraint as
\begin{align}
 \cH_T''&=\cH_T -\frac{\mu^4}{\hat\sigma'}p_{\bn}
 +\frac{\pi V_{\bn}}{\sqrt{h}\kappa^2}p_{\bn}
 +\frac{3\MP^2}{2(\kappa^2 \sigma')^2}h^{ij}\partial_i(\kappa^2 \sigma')
 \partial_j\kappa^2 P_{\bn} \nn
 &\quad -\frac{3\MP^2}{2\kappa^2 \sigma'}h^{ij}\partial_i\kappa^2 D_j
 P_{\bn}
 +\frac{2\MP^2\lambda}{\kappa^2 \sigma'}P_{\bn}
 +\frac{\pi U_{\bn}}{\sqrt{h}\kappa^2}P_{\bn} \approx0,\label{cHT''}\\
 \cH_T&=\frac{2\pi^{ij}\cG_{ijkl}\pi^{kl}}{\sqrt{h}\kappa^2}
 -\frac{1}{2}\sqrt{h}\kappa^2\sR + \sqrt{h}D_iD^i\kappa^2
 +\sqrt{h}\lambda \approx0.
\end{align}
and an extension of the constraint \eqref{cBi} as
\begin{equation}
 \cB_i'=\cB_i-\MP^2\partial_i\left( \frac{P_{\bn}}{\sqrt{h}\sigma'}
 \right) =\partial_i\kappa^2 -\MP^2\partial_i\left(
 \frac{P_{\bn}}{\sqrt{h}\sigma'}  \right) \approx0. \label{cBi'}
\end{equation}
Proving the preservation of $\cH_T$, $\cH_i$, $\cB_i$ and $\cC_i$ is
straightforward, since their structure is rather similar to that in Ref.~\cite{Bufalo:2015wda}.
Note that the Hamiltonian constraint does not depend on $P_{\kappa^2}$
and the constraints $\cH_T$, $\cH_i$, $\cB_i$, $\cC_i$ have vanishing
Poisson brackets with both $P_{\bn}$ and $p_{\bn}$. The constraints
$\cH_T$, $\cH_i$, $\cB_i$, $\cC_i$ satisfy the following Poisson
brackets
(the five omitted Poisson brackets vanish strongly):
\begin{align}
 \pb{\cH_T[\xi],\cH_T[\eta]}&=\int d^3x
 (\xi\partial_i\eta -\eta\partial_i\xi)h^{ij}\bigl[\cH_j
 -\left( P_{\kappa^2} +4\kappa^{-2}\pi^{ij}-\kappa^{-2}h^{ij}\pi \right)
 \cB_j \nn
 &\quad -\partial_jU_{\bn}p_{\bn}-p_\lambda\cC_j
 -\partial_jV_{\bn}p_{\bn} \bigr],\nn
 \pb{\Phi[\chi^i],\cH_T[\xi]}&=\cH_T[\chi^i\partial_i\xi],\nn
 \pb{\Phi[\chi^i],\Phi[\psi^j]}&=\Phi[\chi^j\partial_j\psi^i
 -\psi^j\partial_j\chi^i],\nn
 \pb{\Phi[\chi^i],\cB[\eta^j]}&=\cB[\chi^i\partial_j\eta^j],\nn
 \pb{\Phi[\chi^i],\cC[\eta^j]}&=\cC[\chi^i\partial_j\eta^j].
 \label{scalgebra}
\end{align}
Hence all the constraints are now consistent under time evolution.

The total Hamiltonian \eqref{H3} is a sum of the first class constraints
$\cH_T''$, $\cH_i$, $\pi_N$, $\pi_i$, $P_i$, $p_i$, $\cB_i'$ and
$\cC_i'$. In addition, we have four second class constraints $p_{\bn}$,
$\cC_\lambda$, $P_{\bn}$ and $\Pi$.

\subsection{Counting of physical degrees of freedom}
In order to clarify the nature of the constraint \eqref{cBi} on
$\kappa^2$, we decompose the variables $\kappa^2,P_{\kappa^2}$ as
\begin{equation}\label{kappadec}
\begin{split}
 \kappa^2(t,x)&=\kappa^2_0(t)+\overline{\kappa^2}(t,x),\\
 P_{\kappa^2}(t,x)&=\frac{\sqrt{h}}{\int d^3x\sqrt{h}}
 P_{\kappa^2}^0(t)+\overline{P_{\kappa^2}}(t,x),
\end{split}
\end{equation}
where the zero modes describe the time-dependent average values of
$\kappa^2$ and $P_{\kappa^2}$ over space,
\begin{equation}\label{lambdazero}
 \kappa^2_0(t)=\frac{1}{\int d^3x\sqrt{h}}
 \int d^3x\sqrt{h}\kappa^2(t,x),\quad
 P_{\kappa^2}^0(t)=\int d^3x P_{\kappa^2}(t,x),
\end{equation}
and the oscillating modes have vanishing average values over space,
\begin{equation}\label{noaverage}
 \int d^3x\sqrt{h} \overline{\kappa^2}(t,x)=0,\quad
 \int d^3x\overline{P_{\kappa^2}}(t,x)=0.
\end{equation}
Such a decomposition can always be performed, but for infinite
spaces care must be taken in imposing appropriate (asymptotic) boundary
conditions and in defining the integrals over infinite volumes
(see Refs.\cite{Bufalo:2015wda, Kuchar:1991xd}).
The zero modes satisfy the canonical Poisson bracket
\begin{equation}
 \pb{\kappa^2_0,P_{\kappa^2}^0}=1,
\end{equation}
while the oscillating modes satisfy
\begin{equation}
 \pb{\overline{\kappa^2}(x),\overline{P_{\kappa^2}}(y)}=
 \delta(x-y)-\frac{\sqrt{h}(y)}{\int d^3z\sqrt{h}}
 \equiv \overline{\delta(x,y)},
\end{equation}
where we have defined the overlined $\delta$ function that
satisfies\footnote{Note that
$\overline{\delta(x,y)}\neq\overline{\delta(y,x)}$.}
\[
\int d^3x\,\overline{\delta(x,y)}\sqrt{h}(x)f(x)
=\sqrt{h}(y)\overline{f(y)},\quad
\int d^3y\,\overline{\delta(x,y)}f(y)=\overline{f(x)}.
\]
The Poisson brackets between zero modes and oscillating modes
vanish
\begin{equation}
 \pb{\kappa^2_0,\overline{P_{\kappa^2}}(x)}=0,\quad
 \pb{\overline{\kappa^2}(x),P_{\kappa^2}^0}=0.
\end{equation}
The cosmological constant variables $(\lambda,p_\lambda)$ are decomposed
in the same way to zero modes $(\lambda_0,p_\lambda^0)$ and oscillating
modes $(\overline{\lambda},\overline{p_\lambda})$.
The purpose of the decomposition of the variables $\lambda$
and $\kappa^2$ is to separate the spatially oscillating components
which vanish due to the constraints  \eqref{cBi} and \eqref{cCi}.
When the variables $(\kappa^2,P_{\kappa^2})$ and $(\lambda,p_\lambda)$
are decomposed, the constraints \eqref{cBi} and \eqref{cCi} can indeed
be replaced with local constraints
\begin{equation}\label{ocBocC}
 \overline{\cB}=\overline{\kappa^2}\approx0,\quad
 \overline{\cC}=\overline{\lambda}\approx0,
\end{equation}
since $\partial_i\kappa^2=\partial_i\overline{\kappa^2}=0$ implies that
$\overline{\kappa^2}$ is constant over space and the zero-average
condition
\eqref{noaverage} requires the constant to be zero. The corresponding
first class constraints \eqref{cCi'} and \eqref{cBi'} are replaced with
\begin{align}
 \overline{\cB}'&=\overline{\kappa^2}-\MP^2\overline{\left(
 \frac{P_{\bn}}{\sqrt{h}\sigma'} \right)} \approx0,\label{barcB'}\\
 \overline{\cC}'&=\overline{\lambda}-\mu^4\overline{\left(
\frac{p_{\bn}}{\sqrt{h}\hat\sigma'} \right)} \approx0,\label{barcC'}
\end{align}
where the overline denotes a component whose integral over space
vanishes.

The number of physical degrees of freedom is readily obtained via
Dirac's formula. There are two propagating physical degrees of freedom
for the graviton and two zero modes. The zero modes are the
gravitational and cosmological constants, which do not evolve in time
since the Hamiltonian does not depend on the corresponding canonical
momenta.

\subsection{Elimination of auxiliary variables and classical equivalence
to GR}
\label{secCanVES.re}
Gauge-fixing conditions associated with the generators \eqref{barcB'}
and \eqref{barcC'} can be chosen as
\begin{equation}
 \overline{P_{\kappa^2}}\approx0,\quad \overline{p_\lambda}\approx0.
\end{equation}
Furthermore we choose the gauge conditions for the generators
$P_i\approx0$ and $p_i\approx0$ as
\begin{equation}
 U^i\approx0,\quad V^i\approx0
\end{equation}
Then we have a set of second class constraints as
\begin{equation}\label{scc}
 \phi_I=\left[ p_{\bn}, \cC_\lambda, P_{\bn}, \Pi, \overline{\cB}',
\overline{P_{\kappa^2}}, \overline{\cC}', \overline{p_\lambda},
P_i, U^i, p_i, V^i \right],
\end{equation}
where $I=1,\ldots,12$. We shall use these constraints for the
elimination of the variables $V_{\bn}$, $p_{\bn}$, $U_{\bn}$, $P_{\bn}$,
$\overline{\kappa^2}$, $\overline{P_{\kappa^2}}$, $\overline{\lambda}$,
$\overline{p_\lambda}$, $U^i$, $P_i$, $V^i$ and $p_i$.
In order to set the second class constraints to zero strongly, we
replace the Poisson bracket with the Dirac bracket.
The matrix of Poisson brackets for the constraints \eqref{scc},
\begin{equation}\label{C_IJ}
 C_{IJ}(x,y)=\pb{\phi_I(x),\phi_J(y)},
\end{equation}
has the following nonvanishing components with $I<J$
[Poisson brackets that turn out to be proportional to the constraints
\eqref{scc} --- and thus vanish --- are omitted as well]
\begin{align}
%  C_{11}(x,y)&=0,\nn
 C_{12}(x,y)&=\pb{p_{\bn}(x),\cC_\lambda(y)}
 =\sqrt{h}\frac{\hat\sigma'}{\mu^4}(y)\delta(x-y),\nn
%  C_{13}(x,y)&=0,\nn
%  C_{14}(x,y)&=0,\nn
%  C_{15}(x,y)&=0,\nn
%  C_{16}(x,y)&=0,\nn
%  C_{17}(x,y)&=0,\nn
%  C_{18}(x,y)&=0,\nn
%  C_{21}(x,y)&=\pb{\cC_\lambda(x),p_{\bn}(y)}
%  =-\sqrt{h}\frac{\hat\sigma'}{\mu^4}(x)\delta(x-y),\nn
%  C_{22}(x,y)&=0,\nn
%  C_{23}(x,y)&=0,\nn
 C_{24}(x,y)&=\pb{\cC_\lambda(x),\Pi(y)}
 =-\frac{3}{2}\sqrt{h}\frac{\hat\sigma'}{\mu^4}V_{\bn}(x)
 \delta(x-y),\nn
%  C_{25}(x,y)&=0,\nn
%  C_{26}(x,y)&=0,\nn
%  C_{27}(x,y)&=\pb{\cC_\lambda(x),\overline{\cC}'(y)}
%  =-\frac{\hat\sigma''}{\sqrt{h}(\hat\sigma')^2}p_{\bn}(y)
%  \delta(x-y) +\frac{1}{\int d^3z\sqrt{h}}
%  \frac{\hat\sigma''}{(\hat\sigma')^2}p_{\bn}(x) \approx0
%  =(p_{\bn}\text{-terms})\approx0,\nn
 C_{28}(x,y)&=\pb{\cC_\lambda(x),\overline{p_\lambda}(y)}
 =-\sqrt{h}\frac{\hat\sigma''}{\mu^8}V_{\bn}(x)
 \overline{\delta(x,y)},\nn
%  C_{31}(x,y)&=0,\nn
%  C_{32}(x,y)&=0,\nn
%  C_{33}(x,y)&=0,\nn
 C_{34}(x,y)&=\pb{P_{\bn}(x),\Pi(y)}
 =-\sqrt{h}\frac{\kappa^2}{\MP^2}\sigma'(y)\delta(x-y)
%  \approx-\sqrt{h}\frac{\kappa^2_0}{\MP^2}\sigma'(y)\delta(x-y)
 ,\nn
%  C_{35}(x,y)&=0,\nn
%  C_{36}(x,y)&=0,\nn
%  C_{37}(x,y)&=0,\nn
%  C_{38}(x,y)&=0,\nn
%  C_{41}(x,y)&=0,\nn
%  C_{42}(x,y)&=\pb{\Pi(x),\cC_\lambda(y)}
%  =\frac{3}{2}\sqrt{h}\frac{\hat\sigma'}{\mu^4}V_{\bn}(y)
%  \delta(x-y),\nn
%  C_{43}(x,y)&=\pb{\Pi(x),P_{\bn}(y)}
%  =\sqrt{h}\frac{\kappa^2}{\MP^2}\sigma'(x)\delta(x-y)
%  \approx\sqrt{h}\frac{\kappa^2_0}{\MP^2}\sigma'(x)\delta(x-y),\nn
%  C_{44}(x,y)&=0,\nn
%  C_{45}(x,y)&=\pb{\Pi(x),\overline{\cB}'(y)}
%  =-\frac{3}{2}\sqrt{h}(x) \frac{\int d^3z\sqrt{h}\kappa^2}{\left(
%  \int d^3z\sqrt{h} \right)^2}
%  +\frac{3}{2}\frac{\sqrt{h}\kappa^2(x)}{\int d^3z\sqrt{h}}
%  + (P_{\bn}\text{-terms})
%  =\frac{3}{2}\frac{\sqrt{h}\,\overline{\kappa^2}(x)}
%  {\int d^3z\sqrt{h}} + (P_{\bn}\text{-terms})\approx0,\nn
 C_{46}(x,y)&=\pb{\Pi(x),\overline{P_{\kappa^2}}(y)}
 =\frac{3}{2}\frac{\sqrt{h}(y)\overline{\delta(y,x)}}{\int d^3z\sqrt{h}}
  P_{\kappa^2}^0 -\overline{\delta(x,y)} P_{\kappa^2}(x) \nn
 &\qquad\qquad\qquad\qquad\quad+\sqrt{h}\left( \frac{\sigma'}{\MP^2}
 +\frac{\kappa^2\sigma''}{\MP^4} \right)(x)U_{\bn}(x)
 \overline{\delta(x,y)}
%  &\qquad\qquad\qquad\qquad\approx
%  \frac{\frac{3}{2}\sqrt{h}(y)\overline{\delta(y,x)} -\sqrt{h}(x)
%  \overline{\delta(x,y)}}{\int d^3z\sqrt{h}} P_{\kappa^2}^0 \nn
%  &\qquad\qquad\qquad\qquad\quad+\sqrt{h}\left( \frac{\sigma'}{\MP^2}
%  +\frac{\kappa^2_0\sigma''}{\MP^4} \right)(x)U_{\bn}(x)
%  \overline{\delta(x,y)},\nn
%  C_{47}(x,y)&=\pb{\Pi(x),\overline{\cC}'(y)}
%  =\frac{3}{2}\frac{\sqrt{h}\,\overline{\lambda}(x)}
%  {\int d^3z\sqrt{h}} + (p_{\bn}\text{-terms})\approx0
 ,\nn
 C_{48}(x,y)&=\pb{\Pi(x),\overline{p_\lambda}(y)}
 =\frac{3}{2}\frac{\sqrt{h}(y)\overline{\delta(y,x)}}{\int d^3z\sqrt{h}}
  p_\lambda^0,\nn
%  C_{51}(x,y)&=0,\nn
%  C_{52}(x,y)&=0,\nn
%  C_{53}(x,y)&=0,\nn
%  C_{54}(x,y)&=\pb{\overline{\cB}'(x),\Pi(y)}
%  =-\frac{3}{2}\frac{\sqrt{h}\,\overline{\kappa^2}(y)}
%  {\int d^3z\sqrt{h}} + (P_{\bn}\text{-terms})\approx0,\nn
%  C_{55}(x,y)&=0,\nn
 C_{56}(x,y)&=\pb{\overline{\cB}'(x),\overline{P_{\kappa^2}}(y)}
 =\overline{\delta(x,y)} + (P_{\bn}\text{-terms})
 \approx\overline{\delta(x,y)},\nn
%  C_{57}(x,y)&=0,\nn
%  C_{58}(x,y)&=0,\nn
%  C_{61}(x,y)&=0,\nn
%  C_{62}(x,y)&=0,\nn
%  C_{63}(x,y)&=0,\nn
%  C_{64}(x,y)&=\pb{\overline{P_{\kappa^2}}(x),\Pi(y)}
%  =-\frac{3}{2}\frac{\sqrt{h}(x)\overline{\delta(x,y)}}
%  {\int d^3z\sqrt{h}}  P_{\kappa^2}^0
%   +\overline{\delta(y,x)} P_{\kappa^2}(y) \nn
%  &\qquad\qquad\qquad\qquad\quad-\sqrt{h}\left( \frac{\sigma'}{\MP^2}
%  +\frac{\kappa^2\sigma''}{\MP^4} \right)(y)U_{\bn}(y)
%  \overline{\delta(y,x)} \nn
%  &\qquad\qquad\qquad\qquad\approx
%  \frac{-\frac{3}{2}\sqrt{h}(x)\overline{\delta(x,y)} +\sqrt{h}(y)
%  \overline{\delta(y,x)}}{\int d^3z\sqrt{h}} P_{\kappa^2}^0 \nn
%  &\qquad\qquad\qquad\qquad\quad-\sqrt{h}\left( \frac{\sigma'}{\MP^2}
%  +\frac{\kappa^2_0\sigma''}{\MP^4} \right)(x)U_{\bn}(y)
%  \overline{\delta(y,x)},\nn
%  C_{65}(x,y)&=\pb{\overline{P_{\kappa^2}}(x),\overline{\cB}'(y)}
%  =-\overline{\delta(y,x)} + (P_{\bn}\text{-terms})
%  \approx\overline{\delta(y,x)},\nn
%  C_{66}(x,y)&=0,\nn
%  C_{67}(x,y)&=0,\nn
%  C_{68}(x,y)&=0,\nn
%  C_{71}(x,y)&=0,\nn
%  C_{72}(x,y)&=(p_{\bn}\text{-terms})\approx0,\nn
%  C_{73}(x,y)&=0,\nn
%  C_{74}(x,y)&=\pb{\overline{\cC}'(x),\Pi(y)}
%  =-\frac{3}{2}\frac{\sqrt{h}\,\overline{\lambda}(y)}
%  {\int d^3z\sqrt{h}} + (p_{\bn}\text{-terms})\approx0,\nn
%  C_{75}(x,y)&=0,\nn
%  C_{76}(x,y)&=0,\nn
%  C_{77}(x,y)&=0,\nn
 C_{78}(x,y)&=\pb{\overline{\cC}'(x),\overline{p_\lambda}(y)}
 =\overline{\delta(x,y)} + (p_{\bn}\text{-terms})
 \approx\overline{\delta(x,y)},\nn
%  C_{81}(x,y)&=0,\nn
%  C_{82}(x,y)&=\pb{\overline{p_\lambda}(x),\cC_\lambda(y)}
%  =\sqrt{h}\frac{\hat\sigma''}{\mu^8}V_{\bn}(y)
%  \overline{\delta(y,x)},\nn
%  C_{83}(x,y)&=0,\nn
%  C_{84}(x,y)&=\pb{\overline{p_\lambda}(x),\Pi(y)}
%  =-\frac{3}{2}\frac{\sqrt{h}(x)\overline{\delta(x,y)}}
%  {\int d^3z\sqrt{h}} p_\lambda^0,\nn
%  C_{85}(x,y)&=0,\nn
%  C_{86}(x,y)&=0,\nn
%  C_{87}(x,y)&=\pb{\overline{p_\lambda}(x),\overline{\cC}'(y)}
%  =-\overline{\delta(y,x)} + (p_{\bn}\text{-terms})
%  \approx-\overline{\delta(y,x)},
%  C_{88}(x,y)&=0,\nn
 C_{9,10}(x,y)&=-\delta(x-y),\nn
 C_{11,12}(x,y)&=-\delta(x-y),
\end{align}
where $\sigma''$ and $\hat\sigma''$ are the second derivatives the
scale functions $\sigma$ and $\hat\sigma$. In the components of
Eq.~\eqref{C_IJ} with $I>J$ the coordinates $(x,y)$ are interchanged, e.g.
\begin{equation}
 C_{82}(x,y)=\pb{\overline{p_\lambda}(x),\cC_\lambda(y)}
 =\sqrt{h}\frac{\hat\sigma''}{\mu^8}V_{\bn}(y)
 \overline{\delta(y,x)}.
\end{equation}
Also notice that when the constraints \eqref{scc} are imposed strongly,
the arguments of the scale functions and their derivatives now involve
only the zero modes of $\kappa^2$ and $\lambda$:
\begin{equation}
 \sigma\left( \frac{\kappa^2}{\MP^2} \right)=
 \sigma\left( \frac{\kappa^2_0}{\MP^2} \right),\quad
  \hat\sigma\left( \frac{\lambda}{\mu^4} \right)=
  \hat\sigma\left( \frac{\lambda_0}{\mu^4} \right), \quad\text{etc.}
\end{equation}
The inverse matrix $C^{-1}_{IJ}(x,y)$ is defined by
\begin{equation}
\begin{split}
 \sum_{J=1}^{12}\int d^3yC^{-1}_{IJ}(x,y)C_{JK}(y,z)
 &=\delta_{IK}\delta(x-z),\\
 \sum_{J=1}^{12}\int d^3yC_{IJ}(x,y)C^{-1}_{JK}(y,z)
 &=\delta_{IK}\delta(x-z).
\end{split}
\end{equation}
The Dirac bracket is defined as
\begin{equation}
 \pb{f_1,f_2}_\mathrm{D}=\pb{f_1,f_2}-\sum_{I,J=1}^{12}\int d^3xd^3y
 \pb{f_1,\phi_I(x)}C^{-1}_{IJ}(x,y)\pb{\phi_J(y),f_2},
\end{equation}
where $f_1$ and $f_2$ are any functions or functionals of the canonical
variables. Since the nonvanishing components of the matrix
$C^{-1}_{IJ}(x,y)$ are the components with the indices $\{IJ\}=\{12\}$,
$\{13\}$, $\{17\}$, $\{34\}$, $\{35\}$, $\{37\}$, $\{56\}$, $\{78\}$,
$\{9,10\}$, $\{11,12\}$ and the corresponding components with $I<J$, we
see that the Dirac bracket is equal to the Poisson bracket,
\begin{equation}
 \pb{f_1,f_2}_\mathrm{D}=\pb{f_1,f_2}
\end{equation}
for any $f_1$ and $f_2$ that depend on the remaining canonical
variables $N$, $N^i$, $h_{ij}$, $\kappa^2_0$, $\lambda_0$, $\pi_N$,
$\pi_i$, $\pi^{ij}$, $P_{\kappa^2}^0$ and $p_\lambda^0$.
The Hamiltonian reduces to the GR form
\begin{equation}
 H=\int d^3x\left( N\cH_T+N^i\cH_i +u_N\pi_N+u_N^i\pi_i \right),
\end{equation}
where
\begin{align}
 \cH_T&=\frac{2\pi^{ij}\cG_{ijkl}\pi^{kl}}{\kappa^2_0\sqrt{h}}
 -\frac{\kappa^2_0}{2}\sqrt{h}\sR +\sqrt{h} \lambda_0,\label{cHT.re}\\
 \cH_i&=-2h_{ij}D_k\pi^{jk}.\label{cHi.re}
\end{align}
The gravitational and cosmological constants $\kappa^2_0$ and
$\lambda_0$ depend on time formally, but they do not evolve since
$\cH_T$ is independent of $P_{\kappa^2}^0$ and $p_\lambda^0$.

\subsection{Gauge-fixed action for quantization}
\label{secCanVES.gf}

Next we construct the path integral and the gauge-fixed action for the BRST
formalism. Since the constraints \eqref{cCi'} and \eqref{cBi'} are total
derivatives and their integrals vanish, we have linearly dependent
generators. The quantization of a gauge system with linearly dependent
generators is achieved in the formalism of Ref.~\cite{Batalin:1984jr}.
Fortunately, the situation with the nonlocally linearly dependent
generator associated with the (cosmological constant) variable $\lambda$
is similar to the case of unimodular gravity, which has been described
in Ref.~\cite{Bufalo:2015wda}. The generator associated with the (bare
gravitational constant) variable $\kappa^2$ can be treated in a similar
way in the formalism of Ref.~\cite{Batalin:1984jr}.

First we solve the second-class constraints $(p_{\bn}, \cC_\lambda,
P_{\bn}, \Pi)$ and eliminate the variables $V_{\bn}$, $p_{\bn}$,
$U_{\bn}$ and $P_{\bn}$. As shown in the previous subsection, the Dirac
bracket is equal to the Poisson bracket for the remaining variables.
The generators are denoted by
\begin{equation}\label{G.VES}
 G_\alpha=\left[ \pi_N,\pi_i,\cH_T,\cH_i,\overline{\cB},\overline{\cC},
 P_i,p_i \right]
\end{equation}
and their nonvanishing Poisson/Dirac brackets are given in
Eq.~\eqref{scalgebra}. Here the Hamiltonian and momentum constraints are
defined in Eqs.~\eqref{cHi} and \eqref{cHT2}. The gauge conditions are
written as
\begin{equation}\label{chi.VES}
 \chi^\alpha=\left[ \sigma^0_N,\sigma^i_N,\chi^0,\chi^i,
 \overline{P_{\kappa^2}},\overline{p_{\lambda}},U^i,V^i \right],
\end{equation}
where $\sigma^\mu_N$ fix the lapse and shift functions, $\chi^\mu$
are coordinate conditions for the metric, and the conditions
$\overline{P_{\kappa^2}}$, $\overline{p_{\lambda}}$, $U^i$ and $V^i$
are chosen for simplicity.
Note that the gauge conditions for the generators \eqref{ocBocC} have to
be degenerate to the same degree as the generators; in this case, the
integrals of the conditions over space are fixed.

When the generators $G_\alpha$ are linearly dependent, there
exist right zero eigenvectors $Z^\alpha_a$,
\begin{equation}
 G_\alpha Z^\alpha_a=0.
\end{equation}
The condensed index $\alpha$ labels each local generator at every
point on the spatial hypersurfaces. Summing over such an index
involves an integration over space in addition to a sum over the
components. The latin index $a$ labels the zero eigenvectors.
The vectors $Z^\alpha_a$ are assumed to be linearly independent, i.e.,
we consider a first-stage reducible theory.
The gauge conditions $\chi^\alpha$ have to be similarly dependent as
the generators, so that there exist left zero eigenvectors
$\hat{Z}_\alpha^a$,
\begin{equation}
 \hat{Z}_\alpha^a\chi^\alpha=0.
\end{equation}
The eigenvectors $Z^\alpha_a$ and $\hat{Z}_\alpha^a$ are the right and
left zero vectors of the degenerate Faddeev-Popov operator,
\begin{equation}
 Q^\alpha_{\ \beta}=\pb{\chi^\alpha,G_\beta},
\end{equation}
respectively. In this case, we have the two right
eigenvectors\footnote{The components of these eigenvectors match
those in \eqref{G.VES}, so that for the generators with an index,
$i=1,2,3$, the corresponding component of the eigenvector is understood
to be repeated three times.}
\begin{equation}
\begin{split}
 Z_1^\alpha &=\left[0,0,0,0, \frac{\sqrt{h}}{\int d^3x\sqrt{h}},0,
 0,0\right],\\
 Z_2^\alpha &=\left[0,0,0,0,0, \frac{\sqrt{h}}{\int d^3x\sqrt{h}},
 0,0\right],
\end{split}
\end{equation}
and the left eigenvectors
\begin{equation}
\begin{split}
 \hat{Z}^1_\alpha &=\left[0,0,0,0,1,0,0,0\right],\\
 \hat{Z}^2_\alpha &=\left[0,0,0,0,0,1,0,0\right].
\end{split}
\end{equation}
Hence the Faddeev-Popov ghosts $c^\alpha$, $b_\alpha$ become gauge
fields that require additional gauge fixing. For that purpose the set of
ghosts and Lagrange multipliers $(c^\alpha$, $b_\alpha$, $ \eta_\alpha)$
is extended to
%\cite{Batalin:1984jr}
\begin{equation}
 \Phi_\mathrm{g}=\left( c^\alpha, b_\alpha, \eta_\alpha, C^a,
 B_a, E^a, \theta_a, \vartheta^a \right),
\end{equation}
where $c^\alpha$, $b_\alpha$, $\theta_a$, $\vartheta^a$ are
Grassmann anticommuting variables and the rest are commuting variables.
The path integral and the corresponding effective gauge-fixed action are
written as
\begin{equation}\label{Zgld}
\begin{split}
 Z&=\int\cD q^A\cD p_A\cD\Phi_\mathrm{g}\exp\bigl[
 i \left( S+S_\mathrm{gh+gf} \right) \bigr],\\
 S_\mathrm{gh+gf}&=-\int dt \big[
 b_\alpha Q^\alpha_{\ \beta}c^\beta
 +B_a(\omega^a_\alpha Z^\alpha_b)C^b
 +\eta_\alpha(\chi^\alpha+\sigma^\alpha_aE^a) \\
 &\quad +\theta_a\omega^a_\alpha c^\alpha
 +b_\alpha\sigma^\alpha_a\vartheta^a \big],
\end{split}
\end{equation}
where $q^A$ and $p_A$ denote all the gauge fields and their
canonically conjugated momenta, and $S$ is the action without gauge
fixing. The extra Lagrange multipliers $(\theta_a$, $\vartheta^a)$
impose the gauge conditions $\omega^a_\alpha c^\alpha=0$ and
$b_\alpha\sigma^\alpha_a=0$ on the Faddeev-Popov ghosts, where the
gauge parameters $(\omega^a_\alpha$, $\sigma^\alpha_a)$ are arbitrary.
The variables $B_a$ and $C^a$ are the ghosts for the Faddeev-Popov
ghost fields. The so-called extra ghosts $E^a$ regulate divergent
factors $\delta(0)$ that appear in the original gauge fixing
$\delta(\chi^\alpha)$ with a redundant set of gauge conditions
\eqref{chi.VES}. In our case, we can choose the gauge-fixing parameters
for the
ghosts as
\begin{align}\label{gfparam}
 \omega^1_\alpha&=\left[0,0,0,0,-1,0,0,0\right],\nn
 \omega^2_\alpha&=\left[0,0,0,0,0,-1,0,0\right],\nn
 \sigma_1^\alpha&=\left[0,0,0,0,\frac{\sqrt{h}}{\int d^3x\sqrt{h}},0,
 0,0\right],\nn
 \sigma_2^\alpha&=\left[0,0,0,0,0,\frac{\sqrt{h}}{\int d^3x\sqrt{h}},
 0,0\right],
\end{align}

Integration over the ghost sector gives the path integral as
\begin{equation}\label{Zgld2}
 Z=\int\cD q^A\cD p_A
 \frac{\det\cF^\alpha_{\ \beta}}{\det q^a_b \det\hat{q}^a_b}
 \int\cD E^a\delta(\chi^\alpha+\sigma^\alpha_aE^a) (\det \hat{q}^a_b)
 \exp\left(iS\right),
\end{equation}
where the gauge-fixed Faddeev-Popov operator is defined as
\begin{equation}\label{FPop.gf}
 \cF^\alpha_{\ \beta}=Q^\alpha_{\ \beta}+\sigma^\alpha_a
 \omega^a_\beta,
\end{equation}
and the following matrices are introduced
\begin{equation}\label{qmatrices}
 q^a_b=\omega^a_\alpha Z^\alpha_b,\quad
 \hat{q}^a_b=\hat{Z}^a_\alpha\sigma^\alpha_b.
\end{equation}
The path integral \eqref{Zgld2} is independent of the chosen gauge
parameters $(\omega^a_\alpha$, $\sigma^\alpha_a)$, since both the
ratio of determinants $(\det\cF^\alpha_{\ \beta}/\det q^a_b
\det\hat{q}^a_b)$ and the regulated gauge-fixing factor are
invariant under a change of the gauge parameters.

Since in the present case we chose the gauge fixing of the ghosts so
that
\begin{equation}
 q^a_b=\begin{pmatrix} -1&0\\ 0&-1 \end{pmatrix},\quad
 \hat{q}^a_b=\begin{pmatrix}1&0\\ 0&1 \end{pmatrix},
\end{equation}
the gauge fixing and ghost action for the path integral can be
written in a simpler form without the additional ghosts and Lagrange
multipliers as
\begin{equation}
 S_\mathrm{gh+gf}=-\int dt \big[ b_\alpha\cF^\alpha_{\ \beta}c^\beta
 +\eta_\alpha(\chi^\alpha+\sigma^\alpha_aE^a) \big].
\end{equation}
Furthermore we can trivially integrate over the parts of the ghost
sector that correspond to the generators $\pi_N$, $\pi_i$, $P_i$ and
$p_i$, absorbing them into the normalization of the path integral, so
that the the gauge fixing and ghost action is written as
\begin{multline}\label{SVESgfgh}
 S_\mathrm{gh+gf}= -\int dt \Biggl( b_\mu Q^\mu_{\ \nu}c^\nu
 +\bar{b}_m\bar{Q}^m_{\ \mu}c^\mu
 +\bar{b}_m\bar{\cF}^m_{\ n}\bar{c}^n +\eta_\mu\chi^\mu
 +\bar{\eta}_m\bar{\chi}^m \\
 +\frac{\int d^3x\sqrt{h}\bar{\eta}_m} {\int d^3x\sqrt{h}}E^m \Biggr),
\end{multline}
where we denote
\begin{align}
 \bar{\chi}^m&=\left[ \overline{P_{\kappa^2}}, \overline{p_\lambda}
 \right],\label{barchi}\\
 \bar{G}_m&=\left[ \overline{\cB},\overline{\cC} \right],\label{barG}
\end{align}
and we have defined
\begin{align}
 Q^\mu_{\ \nu}&=\pb{\chi^\mu,\cH_\nu},\quad \cH_\nu=(\cH_T,\cH_i),
 \label{Q.FP}\\
 \bar{Q}^m_{\ \mu}&=\pb{\bar{\chi}^m,\cH_\mu},\label{barQ}\\
 \bar{\cF}^m_{\ n}&=\pb{\bar{\chi}^m(x),\bar{G}_n(y)}
 -\delta^m_{\ n}\frac{\sqrt{h}(x)}{\int d^3z\sqrt{h}}
 =-\delta^m_{\ n}\delta(x-y).
\end{align}
Summing over the repeated indices in Eq.~\eqref{SVESgfgh}, $\mu,\nu=0,1,2,3$
and $m,n=1,2$, includes integration over space. The ghosts $b_\mu,c^\mu$
are associated with diffeomorphisms and the ghosts $\bar{b}_m,\bar{c}^m$
with the generators $\overline{\cB}$ and $\overline{\cC}$.
Integration over the extra ghost $E^m$ imposes $\bar{\eta}_m$ to have
a vanishing zero mode. We obtain the operator \eqref{barQ} for the gauge
conditions \eqref{barchi} as
\begin{align}
 \bar{Q}^1_{\ 0}(x,y)%&=\pb{\overline{P_{\kappa^2}}(x),\cH_T(y)}\nn
 &=\overline{\delta(y,x)} \left(\frac{2\pi^{ij}\cG_{ijkl}\pi^{kl}}
 {\sqrt{h}(\kappa^2)^2} +\frac{1}{2}\sqrt{h}\sR \right)(y) \nn
 &\quad -\overline{\delta(y,x)} \frac{\partial}{\partial y^i}\left(
 \sqrt{h}(y)h^{ij}(y)\frac{\partial}{\partial y^j} \right)\nn
 &\quad +\overline{\delta(y,x)} \frac{\pi(y)P_{\kappa^2}^0}
 {\kappa^2(y) \int d^3z\sqrt{h}},\label{barQ10}\\
 \bar{Q}^1_{\ i}(x,y)%&=\pb{\overline{P_{\kappa^2}}(x),\cH_i(y)}\nn
 &=\overline{\delta(y,x)}\left( \partial_iP_{\kappa^2}(y)
 +P_{\kappa^2}(y)\frac{\partial}{\partial y^i} \right) \nn
 &\quad -\overline{\delta(y,x)}\frac{P_{\kappa^2}^0}
 {\int d^3z\sqrt{h}}\left( \partial_i\sqrt{h}(y)
 +\sqrt{h}(y)\frac{\partial}{\partial y^i} \right)\\
 \bar{Q}^2_{\ 0}(x,y)%&=\pb{\overline{p_\lambda}(x),\cH_T(y)}\nn
 &=-\overline{\delta(y,x)}\sqrt{h}(y) +\overline{\delta(y,x)}
 \frac{\pi(y)p_\lambda^0}{\kappa^2(y) \int d^3z\sqrt{h}},\\
 \bar{Q}^2_{\ i}(x,y)%&=\pb{\overline{p_\lambda}(x),\cH_i(y)}\nn
 &=\overline{\delta(y,x)}\left( \partial_ip_\lambda(y)
 +p_\lambda(y)\frac{\partial}{\partial y^i} \right) \nn
 &\quad -\overline{\delta(y,x)}\frac{p_\lambda^0}
 {\int d^3z\sqrt{h}}\left( \partial_i\sqrt{h}(y)
 +\sqrt{h}(y)\frac{\partial}{\partial y^i} \right).\label{barQ2i}
\end{align}
This completes the calculation of the full gravitational (gauge-fixed)
action for quantization. The action $S+S_\mathrm{gh+gf}$ with
Eq.~\eqref{SVESgfgh} admits the BRST symmetry associated with the
generators $(\cH_\mu,\overline{\cB},\overline{\cC})$ and the gauge
conditions $(\chi^\mu,\overline{P_{\kappa^2}}, \overline{p_\lambda})$.

For the chosen gauge \eqref{barchi}, however, we observe that the ghost
structure associated with the generators \eqref{barG} is essentially
trivial. Integration over the ghosts $\bar{b}_m$ and $\bar{c}^m$
($m=1,2$) indeed gives a unit contribution to the path integral,
$\det\bar{\cF}^m_{\ n}=1$, since the term involving $\bar{b}_m$ and
$c^\mu$ does not contribute to the result due to the lack of a term
involving $b_ \mu$ and $\bar{c}^m$. In other words, here the
functional determinant of Eq.~\eqref{FPop.gf} factors as $\det\cF^\alpha_{\
\beta}=\det Q^\mu_{\ \nu}\times\det\bar{\cF}^m_{\ n}$. Then integration
over the oscillating modes $\overline{\kappa^2}$,
$\overline{P_{\kappa^2}}$, $\overline{\lambda}$ and
$\overline{p_\lambda}$ becomes trivial due to the constraints
\eqref{barchi} and \eqref{barG}. The zero modes $\kappa^2_0$ and
$\lambda_0$ remain. Hence the obtained path integral is the same as the
one obtained for the reduced system in Sect.~\ref{secCanVES.re}. The
path
integral will be discussed further in Sect.~\ref{secPI}.

\section{Path integral and a relation of the gravitational and
cosmological constants}
\label{secPI}
\subsection{Canonical path integral}
The canonical path integral for the gravitational sector of the local
vacuum energy sequestering model is obtained as
\begin{multline}
 Z_\mathrm{VES}=\cN_1\int\prod_{x^\mu}\cD N\cD N^i \cD h_{ij}
 \cD\pi^{ij} \cD\kappa^2_0 \cD P_{\kappa^2}^0 \cD\lambda_0
 \cD p_{\lambda}^0 \delta(\chi^\mu)
 \det\bigl|\pb{\chi^\mu,\cH_\nu}\bigr| \\
 \times\exp\biggl[ i\int dtd^3x\bigl(
 \partial_th_{ij}\pi^{ij} +\partial_t\kappa^2_0P_{\kappa^2}^0
 +\partial_t\lambda_0 p_\lambda^0 -N\cH_T-N^i\cH_i \biggr],
\end{multline}
where $\cN_1$ is a normalization factor and the Hamiltonian and
momentum constraints are given in \eqref{cHT.re} and \eqref{cHi.re}.
The same path integral can either be obtained from the formalism
presented in Sect.~\ref{secCanVES.gf} or it could be written for the
reduced Hamiltonian system obtained in Sect.~\ref{secCanVES.re}.
Matter fields have been excluded for the time being. Integration over
the momenta $P_{\kappa^2}^0$ and $p_\lambda^0$ gives
$\delta(\partial_t\kappa^2_0)\delta(\partial_t\lambda_0)$.
Therefore we decompose $\lambda_{0}$ and $\kappa^{2}_{0}$ to constant
components and oscillating components over time as
\begin{equation}
 \lambda_{0}(t)=\varrho^2\Lambda +\overline{\lambda_{0}}(t),\quad
 \kappa_{0}^{2}(t)=\varrho^2+\overline{\kappa_ {0}^{2}}(t),
\end{equation}
where $\varrho^2$ and $\Lambda$ are gravitational and
cosmological constants, respectively, and the oscillating
components satisfy: $\int dt\overline{\lambda_{0}}=0$,
$\int dt\overline{\kappa_{0}^{2}}=0$. Integration over the momentum
$\pi^{ij}$ can be performed in the standard way (see, e.g., Ref.~\cite{Bufalo:2015wda}).
Assuming that the path integral represents a vacuum transition
amplitude for a vacuum state $\ket{\varrho^2,\Lambda}$ that corresponds
to certain values of $\varrho^2$ and $\Lambda$, we obtain the path
integral as
\begin{multline}\label{ZVES.can}
 Z_\mathrm{VES}(\varrho^2,\Lambda)
%  \equiv\braket{\varrho^2,\Lambda|\varrho^2,\Lambda}
 =\cN_2\int\prod_{x^\mu}\cD g_{\mu\nu}
 g^{00}(-g)^{-\frac{3}{2}} \delta(\chi^{\mu})
 N\varrho^6 \det\left|\pb{\chi^{\mu},\cH_{\nu}}_{\pi^{ij}[h]}\right| \\
 \times\exp\biggl[ i\frac{\varrho^2}{2}\int d^4x\sqrt{-g}
 \left( K_{ij}\cG^{ijkl}K_{kl}+\sR-2\Lambda \right) \biggr],
\end{multline}
where the boundary conditions of the path integral are chosen to be
consistent with the given values of $\varrho^2$ and $\Lambda$.

The extra factor $\varrho^6$ in the measure of Eq.~\eqref{ZVES.can} has not
been absorbed into the normalization factor, since the value of
$\varrho^2$ is set by the boundary conditions of the path integral.
Furthermore, in the next subsection, we will consider a superposition
of states with different values of the gravitational and cosmological
constants, which results in an additional integration over $\kappa^2$
and $\Lambda$, where any additional dependence on those variables has
to be taken into account. Alternatively, instead of including the
factor $\varrho^6$ in the measure, it could be included in the
operator \eqref{Q.FP}. In the Dirac gauge, defined by
$\chi^0=h_{ij}\pi^{ij}$ and $\chi^i=\partial_j(h^{\frac{1}{3}}h^{ij})$,
all the ghosts could be made to carry the same dimension, namely, the
mass dimension $[b_\mu]=[c^\mu]=0$, by including $\varrho^2$ in each
$Q^i_{\ \mu}$, $i=1,2,3$, so that every component of the operator would
have the mass dimension $[Q^\mu_{\ \nu}]=4$. However, we shall keep the
factor in the measure and consider a covariant gauge instead.

Finally, we can transform to a covariant gauge, include matter
(for simplicity without additional gauge symmetries below), and define
the generating functional by including external sources $J^{\mu\nu}$ and
$J_\Psi$ for the metric and the matter fields $\Psi$, respectively,
\begin{multline}\label{ZVES.cov}
 Z_\mathrm{VES}(\varrho^2,\Lambda)[J]
 =\cN_2\int\prod_{x^\mu}\cD g_{\mu\nu}\cD\eta_\rho \cD b_\sigma
 \cD c^\sigma \cD\Psi g^{00}(-g)^{-\frac{3}{2}} \varrho^6\\
  \times\exp\bigg[ i\int d^4x\bigg(
 \frac{\varrho^2}{2}\sqrt{-g}(R-2\Lambda)
 -\eta_\mu\chi^\mu -b_\mu Q^\mu_{\ \nu}c^\nu \\
 +\sqrt{-g}\cL_\mathrm{m}(g_{\mu\nu},\Psi)
 +g_{\mu\nu}J^{\mu\nu} +\Psi J_\Psi \bigg) \bigg],
\end{multline}
where we can use any covariant gauge such as, for example, the harmonic gauge
\begin{align}\label{harmonicgauge}
 \chi^\mu&=\partial_\nu\left( \sqrt{-g}g^{\mu\nu} \right)\approx0,\\
 Q^\mu_{\ \nu}c^\nu&=\partial_\nu\left(
 \partial_\rho\left( \sqrt{-g}g^{\mu\nu}c^\rho \right)
 -\sqrt{-g}g^{\mu\rho}\partial_\rho c^\nu
 -\sqrt{-g}g^{\rho\nu}\partial_\rho c^\mu \right).
\end{align}

\subsection{Relation of the gravitational constant, the
cosmological constant, and the energy density and pressure of matter}
Consider a vacuum state of the universe that is a superposition of
states corresponding to different values of the gravitational and
cosmological constants:
\begin{equation}\label{vacuumO}
 \ket{\Omega}=\int d\varrho^2d\Lambda\,\omega(\varrho^2,\Lambda)
 \ket{\varrho^2,\Lambda}.
\end{equation}
Now the general path integral representation of a vacuum transition
amplitude can be written as
\begin{equation}\label{ZVES}
 Z_\mathrm{VES}\equiv\braket{\Omega|\Omega}
 =\int d\mu(\varrho^2,\Lambda)Z_\mathrm{VES}(\varrho^2,\Lambda),
\end{equation}
where the measure is defined by
\begin{equation}\label{measureGL}
 d\mu(\varrho^2,\Lambda)=\left|\omega(\varrho^2,\Lambda)\right|^2
 d\varrho^2 d\Lambda,
\end{equation}
and we assume
\begin{equation}
 \braket{\varrho^2,\Lambda|\varrho^{2'},\Lambda'}=0
 \quad \mathrm{if}\ \Lambda\neq\Lambda'
 \ \mathrm{or}\ \varrho^2\neq\varrho^{2'}.
\end{equation}
A priori we do not know the measure \eqref{measureGL} for the
integration of the gravitational and cosmological constants. Therefore
we assume that the measure is smooth and includes all values of
$\varrho^2\ge0$ and $\Lambda$.
The generating functional is obtained as
\begin{equation}\label{ZVESmatterJ}
 Z_\mathrm{VES}[J]=\int d\mu(\varrho^2,\Lambda)
 Z_\mathrm{VES}(\varrho^2,\Lambda)[J].
\end{equation}
This differs from the generating functional of fully
diffeomorphism-invariant unimodular gravity \cite{Bufalo:2015wda} in
two ways: we have an additional integration over the gravitational
constant $\varrho^2$ and there is an additional factor $\varrho^6$ in
the measure. Note that the latter can be absorbed into the measure
\eqref{measureGL} of the integral \eqref{ZVESmatterJ}.

The question of much interest is whether we can derive a relation between the
expectation values of $\varrho^2$, $\Lambda$ and the
integrated matter energy density over spacetime. That would be a
generalization of the result obtained in unimodular gravity
\cite{Ng:1990rw,Ng:1990xz,Smolin:2009ti,Bufalo:2015wda}.
This can be done in a semiclassical approximation of the background field
approach to the quantization of a gravitational field theory (for a review
of the background field method see, e.g., Ref.~\cite{Adler:1982ri}).
The idea is that when we integrate over the metric and the matter
fields, the dominant contribution comes from the configurations
$(g_{\mu\nu},\Psi)$ that solve the Einstein field equation
($G_{\mu\nu}+\Lambda g_{\mu\nu}=\varrho^{-2}T_{\mu\nu}$) for given
$\varrho^2$ and $\Lambda$. Then the semiclassical approximation of the
path integral \eqref{ZVES} becomes a sum over such configurations
$(g_{\mu\nu},\Psi)$:
\begin{equation}\label{ZVESsca}
 Z_\mathrm{VES}\approx\int d\mu(\varrho^2,\Lambda)
 \sum_{(g_{\mu\nu},\Psi)} \exp\bigg[ i\int d^4x\sqrt{-g}
 \biggl( \varrho^2\Lambda -\frac{T}{2}+\cL_\mathrm{m} \biggr) \biggr],
\end{equation}
where the trace of the Einstein equation was used to write
$R=4\Lambda-\varrho^{-2}T$. Here $\varrho^2$ and $\Lambda$ are the
renormalized gravitational and cosmological constants. According to the
stationary phase approximation the integral \eqref{ZVESsca} is dominated
by solutions for which the on-shell action vanishes.
For a perfect fluid the on-shell action is given by the integral
of pressure $p$ over spacetime \cite{Brown:1992kc},
$ \int d^4x\sqrt{-g}\cL_\mathrm{m}=\int d^4x\sqrt{-g}p$,
and the trace of the energy-momentum tensor is $T=-\rho+3p$, where
$\rho$ is the energy density. If several perfect-fluid components are
considered, we have the total pressure $p=\sum_ap_a$ and the total
energy density $\rho=\sum_a\rho_a$ instead.
Hence we see that the most likely values of the gravitational constant
and the cosmological constant are related to the average values of the
total pressure and the total energy density over the whole spacetime as
\begin{equation}\label{varrhoLambda}
 \varrho^2\Lambda%=\frac{\Lambda}{8\pi G}
 \approx\frac{\int d^4x\sqrt{-g}\left( \frac{T}{2}
 -\cL_\mathrm{m} \right)}{\int d^4x\sqrt{-g}}
 =\frac{\int d^4x\sqrt{-g}(p-\rho)}{2\int d^4x\sqrt{-g}}
 =\frac{1}{2}\langle p-\rho\rangle.
\end{equation}
This relation approximates the product $\varrho^2\Lambda$ (or the ratio
$\Lambda/G$) of the gravitational constant and the cosmological
constant, but it does not tell us anything about their separate values.
Hence the relation \eqref{varrhoLambda} has a very different
nature compared to that obtained in unimodular gravity, although the
form of the relation is similar.

Naturally, estimating the average values of the total pressure and the
total energy density over the history of the universe is quite
difficult, which gives a reason to doubt the usefulness of the relation
\eqref{varrhoLambda}. Still it is interesting that the local formulation
of vacuum energy sequestering implies such a relation between the two
given fundamental constants of nature.

\section{Topological induced gravity}
\label{secIG}

For linear functions $\sigma$ and $\hat\sigma$ the action of the local
vacuum energy sequestering model \eqref{SVES2} reduces to
\begin{equation}\label{SIG}
 S_\mathrm{l}=\int d^4x\sqrt{-g} \biggl[ \frac{\kappa^2}{2}\left( R
 +\MP^{-2}\nabla_\mu U^\mu \right)
 -\lambda\left( 1-\mu^{-4}\nabla_\mu V^\mu \right)
 +\cL_\mathrm{m}(g_{\mu\nu},\Psi) \biggr],
\end{equation}
where we have chosen $\sigma(s)=\frac{1}{2}s$ and $\hat\sigma(s)=s$ for
simplicity. Incidentally, this action is essentially the same as the
recently discussed  ``topological induced gravity''
\cite{Oda:2016xls}.\footnote{The scales $\MP^{-2}$ and $\mu^{-4}$ could
be absorbed into the vector fields $U^\mu$ and $V^\mu$, respectively,
but that would obscure the relation to the vacuum energy sequestering
theory.} It is thus clear that the so-called topological induced gravity
is the simplest special case of the local vacuum energy sequestering
model. The action differs from the fully diffeomorphism-invariant version
of unimodular gravity \cite{Bufalo:2015wda} by the presence of the vector
$U^\mu$ and a variable gravitational coupling $\kappa^2$.
The action can again be rewritten as in Eq.~\eqref{SVES3}.
The canonical structure is identical to that of local vacuum energy
sequestering obtained in Sect.~\ref{secCanVES} with the
substitutions $\sigma'=\frac{1}{2}$ and $\hat\sigma'=1$.

The action \eqref{SIG} can be written in a BRST-exact form by including
an appropriate ghost action \cite{Oda:2016xls}. This justifies the
label ``topological''. It is natural to ask whether such a BRST-exact
formulation can be generalized to the local theory of vacuum energy
sequestering with nonlinear functions $\sigma$ and $\hat\sigma$. This
will be explored next.

\section{Induced theory of vacuum energy sequestering from gauge fixing}
\label{secIG-BRST}

\subsection{Gravitational action from gauge fixing}
Here we consider a BRST-exact formulation for the local version of
vacuum energy sequestering. The gravitational action \eqref{SVES2} will
be seen to emerge as a gauge-fixing action. The approach is similar to
those of Refs.~\cite{Oda:2016xls} and \cite{Nojiri:2016mlb}.

Consider a theory for two vector fields $U^\mu$ and $V^\mu$ on a
curved spacetime with metric $g_{\mu\nu}$, and optionally some matter
fields $\Psi$. The theory is assumed to be diffeomorphism invariant.
The gravitational action of the theory is assumed to vanish initially,
i.e., we consider an action of the form
\begin{equation}
 S=S_\mathrm{v}[g_{\mu\nu},U^\mu,V^\nu]
 +S_\mathrm{m}[g_{\mu\nu},\Psi].
\end{equation}
Including couplings between matter fields $\Psi$ and the vector fields
$U^\mu$ and $V^\mu$ would be possible as well.
Hence we have no theory of gravity in the beginning. We assume that the
action for the vector fields possesses a gauge symmetry under
transformations of $U^\mu$ and $V^\mu$. In the simplest case, one could
consider a vanishing action, $S_\mathrm{v}=0$, so that the action would
be invariant under any transformation of $U^\mu$ and $V^\mu$. Here we
consider that the action $S_\mathrm{v}$ is invariant under the following
two gauge transformations generated by infinitesimal scalar field
parameters $\alpha$ and $\hat\alpha$ with zero mass dimension:
\begin{equation}\label{UVgt}
 \delta_\alpha U^\mu=\MP^2\nabla^\mu\alpha,\quad
 \delta_{\hat\alpha} V^\mu=\mu^2\nabla^\mu\hat\alpha.
\end{equation}
The gauge-fixing action for $U^\mu$ and $V^\mu$ can be chosen as the
gravitational part of the local vacuum energy sequestering action
\eqref{SVES2},
\begin{equation}\label{Sgf}
 S_\mathrm{gf}=\int d^4x\sqrt{-g}\biggl[ \frac{\kappa^2}{2}R -\lambda
+\sigma\left( \frac{\kappa^2}{\MP^2} \right) \nabla_\mu U^\mu
 +\hat\sigma\left( \frac{\lambda}{\mu^4} \right) \nabla_\mu V^\mu
 \biggr],
\end{equation}
where $\kappa^2$ and $\lambda$ are the auxiliary scalar fields
required for imposing the gauge-fixing conditions for the gauge
symmetry under Eq.~\eqref{UVgt}.
The gauge conditions are written as
\begin{equation}\label{gfc}
 \sigma'\left( \frac{\kappa^2}{\MP^2} \right) \nabla_\mu U^\mu
 +\frac{\MP^2}{2}R=0,\quad
 \hat\sigma'\left( \frac{\lambda}{\mu^4} \right) \nabla_\mu V^\mu
 -\mu^4=0.
\end{equation}
Note that the auxiliary variables $\kappa^2$ and $\lambda$ appear in
the gauge conditions \eqref{gfc} when the functions $\sigma$ and
$\hat\sigma$ are nonlinear, since then the gauge-fixing Lagrangian
\eqref{Sgf} is not linear in the auxiliary fields, which is a rather
uncommon situation.

We have chosen the gauge-fixing action \eqref{Sgf} to specifically
match the action of the local formulation of vacuum energy sequestering.
However, the gauge-fixing action could be chosen in a number of
different ways. There are three kinds of changes that could be
considered. First, the curvature part of the action \eqref{Sgf} could be
changed. For example, the gravitational part of the action could be
defined to include higher-order curvature terms, but that would
require additional coupling constants or fields, unless the coefficients
of the higher-curvature terms are set to $\pm\kappa^2\MP^{-n}$,
$n=2,4,\ldots$, which would give weak couplings that are preferable
regarding long-distance behavior.\footnote{One of the
candidates for a modification of the gravitational sector would be Weyl
gravity, due to its known structure and relevance. However, as we have
discussed in the last paragraph of Sect.~\ref{secIntVESglobal}, the
smallness of the cosmological constant in the sequestering mechanism is
due to two approximate symmetries, namely, scaling and shift symmetries,
which are broken by the gravitational sector. Thus, since pure Weyl
gravity is scale invariant, there would be no gravitational scale
$\kappa^2$, but rather a dimensionless gravitational coupling.
Furthermore, the cosmological constant term could not be included in
Weyl gravity, since it ruins the canonical and geometric structure due
to the appearance of a constraint $\sqrt{-g}=0$ \cite{Kluson:2013hza}.
Extending the action \eqref{Sgf} with the conformally invariant Weyl
action would not imply such problems, since then the approximate
scaling and shift symmetries persist, and the canonical structure of the
Einstein-Hilbert plus Weyl action is known to be consistent
\cite{Kluson:2013hza}.}
In principle, we could even abandon full diffeomorphism invariance and
use, for example, the Lagrangian density of Ho\v{r}ava-Lifshitz gravity
\cite{Horava:2009uw} (with a variable $\kappa^2$) in place of the
Einstein-Hilbert term, $\frac{\kappa^2}{2}\sqrt{-g}R$.
Second, the part of the gauge-fixing action \eqref{Sgf} that involves
$U^\mu$ and $V^\mu$ could be chosen differently, e.g. to include
quadratic terms $U_\mu U^\mu$ or direct coupling to curvature etc. That
kind of modification, however, could lead to a completely different type
of theory, since we would no longer obtain the field equations
$\nabla_\mu \kappa^2=0$ and $\nabla_\mu\lambda=0$; the same happens when
a nonvanishing action $S_\mathrm{v}$ for the vector fields is included.
Finally, we could consider different gauge symmetry transformations
instead of Eq.~\eqref{UVgt}. Above we have considered the simplest gauge
transformations that imply scalar ghost fields. The gauge
transformations could be parametrized by vector fields or tensor fields
of higher rank, which would require vector ghost fields or higher-rank
tensor ghost fields, respectively. In summary, the approach explored
here can be used to construct a wide variety of gravitational models
with variable gravitational couplings and a variable cosmological
(constant) parameter.

In order to obtain a BRST-invariant (and eventually BRST-exact)
action, we introduce Grassman-odd ghost fields $c,\hat{c}$ and antighost
fields $b,\hat{b}$. The two BRST transformations can be obtained from
the gauge transformations \eqref{UVgt}  as
\begin{align}
 \delta_B\kappa^2&=0,\quad \delta_B U^\mu=\epsilon\MP^2
 \nabla^\mu c,\quad \delta_B c=0,\quad \delta_B b=
 \epsilon\sigma\left( \frac{\kappa^2}{\MP^2} \right),\label{BRSTforU}\\
 \hat\delta_B\lambda&=0,\quad \hat\delta_B V^\mu=\hat\epsilon\mu^2
 \nabla^\mu \hat{c},\quad \hat\delta_B \hat{c}=0,\quad
 \hat\delta_B \hat{b}=\hat\epsilon\hat\sigma\left( \frac{\lambda}{\mu^4}
 \right),\label{BRSTforV}
\end{align}
where $\epsilon$ and $\hat\epsilon$ are infinitesimal anticommuting
parameters. Note that the gauge transformations \eqref{UVgt} are Abelian
so that the BRST transformations of the ghosts $(c,\hat{c})$ vanish.
The BRST transformations of the antighosts $(b,\hat{b})$ are nonlinear
in the auxiliary fields due to the corresponding nonlinearity of the
gauge-fixing Lagrangian \eqref{Sgf}. The ghost action can be obtained as
\begin{equation}\label{ghS}
\begin{split}
 S_\mathrm{gh}&=-\int d^4x\sqrt{-g}\left( b\MP^2 \nabla_\mu\nabla^\mu c
 +\hat{b}\mu^2 \nabla_\mu\nabla^\mu\hat{c} \right)\\
 &=\int d^4x\sqrt{-g}\left( \MP^2 \nabla_\mu b\nabla^\mu c
 +\mu^2 \nabla_\mu\hat{b}\nabla^\mu\hat{c} \right)
\end{split}
\end{equation}
The gravitational action, $S_g=S_\mathrm{gf}+S_\mathrm{gh}$,
is BRST invariant under Eqs.~\eqref{BRSTforU} and \eqref{BRSTforV}.
The gravitational part of the action is also BRST exact, since it can
be written as
\begin{equation}\label{BRSTexact}
 S_g=\int d^4x\sqrt{-g} \left( s_B\left[b\left( \frac{\kappa^2}
 {2\sigma\left( \frac{\kappa^2}{\MP^2} \right)} R
 +\nabla_\mu U^\mu \right)\right]
 -\hat{s}_B\left[\hat{b}\left( \frac{\lambda}{\hat\sigma\left(
 \frac{\lambda}{\mu^4} \right)}-\nabla_\mu V^\mu \right)\right]
 \right),
\end{equation}
where $s_B$ and $\hat{s}_B$ are the Slavnov variations corresponding to
Eqs.~\eqref{BRSTforU} and \eqref{BRSTforV}, respectively.
Hence it can be considered as a topological field theory
\cite{Witten:1988ze}.
The full gauge-fixed action is now given as $S=S_g+S_\mathrm{v}
+S_\mathrm{m}$.

We should remark that the BRST-exact formulation of
local vacuum energy sequestering should not be confused with the
gauging procedure followed in Sect.~\ref{secIntVESlocal}. The main
conceptual difference between the approaches is the way the gauge
symmetry is imposed: the local formulation of vacuum energy
sequestering is obtained by gauging the global sequestering mechanism,
while here we assume a further gauge symmetry in order to obtain a
BRST-exact form of the action. The present approach can be seen as an
extension of the local formulation of vacuum energy sequestering.

\subsection{Hamiltonian analysis}
Our goal is to find a canonical formulation of the full gauge-fixed action
$S_g=S_{\mathrm{gf}}+S_{\mathrm{gh}}$, where $S_{\mathrm{gf}}$ is
defined in Eq.~\eqref{Sgf} and $S_{\mathrm{gf}}$ is defined in \eqref{ghS}.
Note that the canonical analysis of $S_{\mathrm{gf}}$ has
been done in the previous section, so that we focus
on the ghost action $S_{\mathrm{gh}}$. To begin with we
rewrite it in the $3+1$ formalism,
\begin{align}
S_\mathrm{gh}=\MP^2\int dt d^3x \sqrt{h}N(-\nabla_n b \nabla_n c+h^{ij}
\partial_i b\partial_jc) \nn
+\mu^2\int dt d^3x\sqrt{h}N( -\nabla_n\hat{b}\nabla_n \hat{c}+
h^{ij}\partial_i b\partial_j c),
\end{align}
so that we have conjugate momenta
\begin{align}
p_b=\frac{\delta \mathcal{L}_{gh}}{\delta^L \partial_t b}=
-\MP^2\sqrt{h}\nabla_n c  , \quad
p_c= \frac{\delta \mathcal{L}_{gh}}{\delta^L \partial_t c}
=\MP^2\sqrt{h}\nabla_n b, \nn
p_{\hat{b}}=\frac{\delta \mathcal{L}_{gh}}{\delta^L \partial_t \hat{b}}=
-\mu^2\sqrt{h}\nabla_n \hat{c}  , \quad
p_{\hat{c}}= \frac{\delta \mathcal{L}_{gh}}{\delta^L \partial_t \hat{c}}
=\mu^2\sqrt{h}\nabla_n \hat{b},
\end{align}
with the following nonvanishing (graded) Poisson brackets:
\begin{align}
 \pb{c(x),p_c(y)}&=\pb{p_c(y),c(x)}-\delta(x-y) , \nn
 \pb{b(x),p_b(y)}&=\pb{p_b(y),b(x)}=-\delta(x-y) , \nn
\pb{\hat{c}(x),p_{\hat{c}}(y)}&=\pb{p_{\hat{c}}(y),\hat{c}(x)}=
-\delta(x-y) , \nn
\pb{\hat{b}(x),p_{\hat{b}}(y)}&=\pb{p_{\hat{b}}(y),\hat{b}(x)}
=-\delta(x-y) .
\end{align}
Then it is easy to find the ghost contributions to the Hamiltonian and
diffeomorphism constraints as
\begin{align}
\cH^\mathrm{gh}_T&=\frac{1}{\MP^2\sqrt{h}}
p_c p_b-M^2_P\sqrt{h}h^{ij}\partial_i b\partial_j c
+ \frac{1}{\mu^2\sqrt{h}} p_{\hat{c}} p_{\hat{b}}
-\mu^2\sqrt{h}h^{ij}\partial_i \hat{b}\partial_j \hat{c},
\nonumber \\
\cH^\mathrm{gh}_i&=\partial_ic p_c+\partial_i
bp_b+\partial_i\hat{c}p_{\hat{c}}+
\partial_i \hat{b} p_{\hat{b}} .
\end{align}
Using the standard Noether method, we derive the conserved BRST
currents as
\begin{align}
J_{BRST}^\mu&=\sqrt{-g}\MP^2\sigma\left(\frac{\kappa^2}{\MP^2}\right)
g^{\mu\nu}\nabla_\nu c,\\
\hat{J}_{BRST}^\mu&=\sqrt{-g}\mu^2\hat{\sigma}\left(\frac{\lambda}
{\mu^2}\right) g^{\mu\nu}\nabla_\nu \hat{c},
\end{align}
and hence we have corresponding conserved charges expressed in terms of
the canonical variables,
\begin{align}
 Q_{BRST}&=\int d^3x \sigma\left(\frac{\kappa^2}{\MP^2}\right)p_b,\\
 \hat{Q}_{BRST}&=\int d^3x\hat{\sigma}
 \left(\frac{\lambda}{\mu^2}\right)p_{\hat{b}}.
\end{align}
Since we want these charges to act nontrivially on $U^\mu$
and $V^\mu$, respectively, we add to them linear combinations of the
constraints $P_{\bn},P_i$ and $p_{\bn},p_i$, respectively.
In fact, note that \eqref{UVgt} implies
\begin{align}
\delta_\alpha U_{\bn}=
%n_\mu \delta U^\mu=n^\mu M_P^2
%\nabla_\mu\alpha=
M_P^2\nabla_n\alpha , \quad
\delta_\alpha U^i
%=(\delta^i_\mu+n^in_\mu)\delta U^\mu=
%M_P^2(\delta^i_\mu+n^i n_\mu)\nabla^\mu \alpha=
M_P^2h^{ij}\partial_j\alpha ,
\end{align}
and also
\begin{equation}
\delta_{\hat{\alpha}}V_{\bn}=\mu^2 \nabla_n \hat{\alpha} ,
\quad
\delta_{\hat{\alpha}}V^i=\mu^2 h^{ij}\nabla_j\hat{\alpha} .
\end{equation}
From these transformation rules we deduce the BRST
transformations
\begin{equation}
\delta_B U_{\bn}=-\epsilon\frac{1}{\sqrt{h}}p_b , \quad
\delta_B U^i=\epsilon M_P^2 h^{ij}\partial_j c,
\end{equation}
and hence we find that the extended BRST operator has the form
\begin{equation}\label{Q_BRST}
Q_{BRST}=\int d^3x \left(
\sigma\left(\frac{\kappa^2}{\MP^2}\right)p_b-\frac{p_b}{\sqrt{h}}
P_{\bn}+M_P^2P_i h^{ij}\partial_j c\right).
\end{equation}
In the same way, we write an extended form of the BRST operator
$\hat{Q}_{BRST}$ as
\begin{equation}\label{hatQ_BRST}
 \hat{Q}_{BRST}=\int d^3x\left(\hat{\sigma}
 \left(\frac{\lambda}{\mu^2}\right)p_{\hat{b}}
 -\frac{p_{\hat{b}}}{\sqrt{h}}
p_{\bn}+\mu^2p_i h^{ij}\partial_j \hat{c}\right).
\end{equation}
The BRST charges have the following nonvanishing Poisson brackets with
the canonical  variables:
\begin{align}
\pb{Q_{BRST},b}&=-\sigma\left(\frac{\kappa^2}{\MP^2}
\right) , \quad
\pb{Q_{BRST},P_{\kappa^2}}=\frac{1}{\MP^2}
\sigma'\left(\frac{\kappa^2}{\MP^2}\right)p_b , \nn
\pb{Q_{BRST},U_{\bn}}&=\frac{1}{\sqrt{h}}p_b , \quad \pb{Q_{BRST},
U^i}=-M_P^2 h^{ij}\partial_j c , \nn
\pb{\hat{Q}_{BRST},\hat{b}}&=-\hat{\sigma}\left(\frac{\lambda}{\mu^2}
\right) , \quad
\pb{\hat{Q}_{BRST},p_\lambda}=
\frac{1}{\mu^2}\hat{\sigma}'
\left(\frac{\lambda}{\mu^2}\right)p_{\hat{b}} , \nn
\pb{\hat{Q}_{BRST},V_{\bn}}&=\frac{1}{\sqrt{h}}p_{\hat{b}} ,
\quad
\pb{\hat{Q}_{BRST},V^i}=-\mu^2 h^{ij}\partial_j\hat{c} .
\label{BRSTcano}
\end{align}
The BRST charges \eqref{Q_BRST} and \eqref{hatQ_BRST} Poisson commute
with the second-class constraints \eqref{cClambda} and \eqref{Pi},
which explains why the momenta $P_{\kappa^2}$ and $p_\lambda$ must have
nonvanishing BRST transformations \eqref{BRSTcano}, while
the conjugated variables $\kappa^2$ and $\lambda$ do not change under
the given BRST transformations.

The Poisson brackets between the BRST charges and the Hamiltonian
constraint are now now easily obtained as
\begin{align}
\pb{Q_{BRST},\cH''_T+\cH_T^{ghost}}&\approx
\sigma'\sqrt{h}h^{ij}\partial_i\kappa^2 \partial_jc
=\sigma'\sqrt{h}h^{ij}\cB_i\partial_jc \approx 0 ,\\
\pb{\hat{Q}_{BRST},\cH''_T+\cH_T^{ghost}}&\approx
\hat{\sigma}'\sqrt{h}h^{ij}\partial_i \lambda \partial_j\hat{c}
=\hat{\sigma}'\sqrt{h}h^{ij}\cC_i\partial_j \hat{c}\approx 0,
\end{align}
up to terms proportional to the constraints $P_{\bn}\approx 0$ and
$p_{\bn}\approx 0$. The BRST charges have vanishing Poisson brackets
with all the other constraints. Then it is easy to derive the following
relation
\begin{equation}
\pb{Q_{BRST},-\frac{b}{\sigma}\cH''_T}
%\cH''_T-b
%\frac{\sigma'}{\sigma}\partial_i(\kappa^2)h^{ij}
%\sqrt{h}\partial_j c
=\cH''_T -b\frac{\sigma'}{\sigma}\cB_i h^{ij}
\sqrt{h}\partial_j c .
\end{equation}
Then we could replace $\cH''_T$ in the definition of the Hamiltonian
with this Poisson bracket when the expression proportional to $\cB_i$ is
absorbed into the corresponding Lagrange multiplier. In other words, we
see that the Hamiltonian has the schematic form
\begin{equation}\label{HBRSTQ}
H=\pb{Q_{BRST},\Psi}+\pb{\hat{Q}_{BRST},\hat{\Psi}} ,
\end{equation}
where the explicit form of the gauge-fixing fermions $\Psi$
and $\hat{\Psi}$ is not important for us. On the other hand, the fact
that the Hamiltonian can be written in the form \eqref{HBRSTQ} is a
consequence of the fact that the gravitational action $S_{\mathrm{g}}$
\eqref{BRSTexact} is BRST exact.

\section{A simple model for including the cosmological constant as a
topological field theory}
\label{secNo}

Let us consider the model proposed in Ref.~\cite{Nojiri:2016mlb}. In this
model, the cosmological constant part of the Lagrangian is BRST exact,
while the Einstein-Hilbert gravitational part appears as usual. In
other words, this model is an extension of unimodular gravity in the
same way that our formulation in Sect.~\ref{secIG-BRST} extends local
vacuum energy sequestering. In the model of Ref.~\cite{Nojiri:2016mlb},
however, only the cosmological constant part of the action is made
topological, while in Sect.~\ref{secIG-BRST} the whole gravitational
action of the local version of vacuum energy sequestering was formulated
as a topological field theory. We should note that the model of
Ref.~\cite{Nojiri:2016mlb} is unlikely to actually solve any cosmological
constant problem, since it lacks a mechanism for ensuring the
perturbative stability of the cosmological constant. In this respect,
the model is on par with conventional unimodular gravity. Nevertheless,
the model is an interesting example for building a topological field
theory for the cosmological constant.

The model is defined by the action
\begin{equation}
S=\int d^4x\sqrt{-g}\left(\frac{\MP^2}{2}R-\lambda
+\frac{1}{\mu^3}\partial_\mu \lambda g^{\mu\nu}\partial_\nu
\phi-\partial_\mu b g^{\mu\nu}\partial_\nu c\right),
\end{equation}
where $\phi$ is a scalar field and the fields $b$ and $c$
are Grassmann-odd ghosts.
The action is invariant under BRST transformation
\begin{equation}
\delta \lambda=\delta c=0 , \quad
\delta \phi=\epsilon c ,\quad \delta b=\frac{1}{\mu^3}\epsilon
\lambda  ,
\end{equation}
where $\epsilon$ is a global fermionic (anticommuting) parameter. Note
that this BRST transformation implies the existence of BRST current
$j^\mu_{BRST}$ in the form
\begin{equation}
j^\mu_{BRST}=\frac{1}{\mu^3}\sqrt{-g}
\left(-cg^{\mu\nu}\partial_\nu \lambda+g^{\mu\nu}\partial_\nu c
\lambda\right) ,\quad  \partial_\mu j^\mu_{BRST}=0 .
\end{equation}
As a result we have a conserved charge
\begin{equation}
Q_{BRST}=\int d^3x j^0=\frac{1}{\mu^3}\int d^3x
\left(-c \sqrt{-g}g^{0\nu}\partial_\nu \lambda
+\sqrt{-g}g^{0\nu}\partial_\nu c\lambda\right).
\end{equation}
Our goal is to proceed to the Hamiltonian formalism. Note that
in the $3+1$ formalism the action has the form
\begin{align}
S=\int d^4x N\sqrt{h}\biggl(\frac{\MP^2}{2}\left(K_{ij}\cG^{ijkl}K_{kl}
+\sR\right)-\lambda-\frac{1}{\mu^3}\nabla_n \lambda\nabla_n\phi
+\frac{1}{\mu^3}\partial_i\lambda h^{ij}\partial_j\phi \nn
+\nabla_n b\nabla_n c -h^{ij}\partial_i b\partial_j c\biggr) .
\end{align}
In the same way as in previous sections we obtain
\begin{equation}
\pi^{ij}=\frac{\MP^2}{2}\cG^{ijkl}K_{kl} , \quad
\pi_N\approx 0  ,\quad  \pi_i\approx 0  ,  \quad
p_\lambda=-\frac{\sqrt{h}}{\mu^3}\nabla_n\phi ,\quad  p_\phi=-
\frac{\sqrt{h}}{\mu^3}\nabla_n\lambda .
\end{equation}
In case of the Grassman-odd variables we have to be careful
with the definition of the momenta. We define them in terms of the
variation from the left as
\begin{equation}
p_b=\sqrt{h}\nabla_n c , \quad
p_c=-\sqrt{h}\nabla_n b ,
\end{equation}
so that the Hamiltonian is equal to
\begin{align}
H&=\int d^3x \left(N\mathcal{H}_T+N^i\mathcal{H}_i\right) , \nn
\mathcal{H}_T&=\frac{2}{\MP^2\sqrt{h}}\pi^{ij}\cG_{ijkl}\pi^{kl}
-\frac{\mu^3}{\sqrt{h}}p_\phi p_\lambda-\frac{1}{\sqrt{h}}
p_c p_b-\nn
&\quad -\frac{\MP^2}{2}\sqrt{h}\sR +\sqrt{h}\lambda
-\frac{1}{\mu^3}\sqrt{h}
\partial_i \lambda h^{ij}\partial_j\phi +\sqrt{h}h^{ij}
\partial_i b\partial_j c , \nn
\mathcal{H}_i&=-2h_{ik}D_i \pi^{ik}+\partial_i c p_c+\partial_i b
p_b+\partial_i\lambda p_\lambda
+\partial_i \phi p_\phi .
\end{align}

The BRST charge takes the form
\begin{equation}
Q_{BRST}=\int d^3x\left(-cp_\phi-\frac{1}{\mu^3}p_b\lambda\right),
\end{equation}
so that
\begin{align}
\pb{Q_{BRST},\phi}&=c , \quad \pb{Q_{BRST},b}=\frac{1}{\mu^3}\lambda ,
\nn
\pb{Q_{BRST},p_c}&=p_\phi , \quad \pb{Q_{BRST},p_\lambda}=
-\frac{1}{\mu^3}p_b ,
\end{align}
using also the fact that $b,p_b$ and $c,p_c$ have graded
Poisson brackets
\begin{align}
\pb{c(x),p_c(y)}=\pb{p_c(y),c(x)}=-\delta(x-y) , \nn
\pb{b(x),p_b(y)}=\pb{p_b(y),b(x)}=-\delta(x-y) .
\end{align}
Then it is easy to see that
\begin{equation}
\pb{Q_{BRST},\cH_T}=0 , \quad
\pb{Q_{BRST},\cH_i}=0 ,
\end{equation}
which is the reflection of the fact that $Q_{BRST}$ is conserved.
The question is whether the existence of this charge
can remove negative unphysical states. Let us consider a
phase-space function $Z$ with Grassmann parity $|Z|$. Then using
the generalized Jacobi identity
\begin{equation}
\pb{X,\pb{Y,Z}}=\pb{\pb{X,Y},Z}+(-1)^{|X||Y|}\pb{Y,\pb{X,Z}}.
\end{equation}
Now since it is a phase-space function its Poisson bracket has to weakly
vanish on the constraint surface
\begin{equation}
\pb{Z,\cC_A}=u_A^{ \ B}\cC_B ,
\end{equation}
where $\cC_A=(\cH_T,\cH_i,\pi_N,\pi_i)$. Note that
since $Z$ has grading $|Z|$ and the constraints have $|\cC_A|=0$,
$u_A^{\ B}$ has to have the same grading as $Z$.
Let us use the graded Jacobi identity above for $X=Q_{BRST}$ and
$Y=\cC_A$ as
\begin{equation}
\pb{Q_{BRST},\pb{\cC_A,Z}}=
\pb{\pb{Q_{BRST},\cC_A},Z}+\pb{\cC_A,
\pb{Q_{BRST},Z}}.
\end{equation}
Using the fact that the first expression on the right-hand
side is equal to zero, we obtain
\begin{equation}
\pb{Q_{BRST},u_A^{ \ B}\cC_B}=\pb{\cC_A,\pb{Q_{BRST},Z}}.
\end{equation}
Since $\pb{Q_{BRST},\cC_A}=0$, we obtain
\begin{equation}
\pb{\cC_A,\pb{Q_{BRST},Z}}=0 ,
\end{equation}
which implies that
\begin{equation}
\pb{Q_{BRST},Z}=v^A\cC_A .
\end{equation}
In other words, whenever $Z$ is a physical variable that is
invariant under diffeomorphism, it has to have a weakly vanishing
Poisson bracket with $Q_{BRST}$.

\section{Conclusions}
\label{conc}

We have studied the Hamiltonian formalism and path integral
quantization of the local version of vacuum energy sequestering.
The path integral \eqref{ZVES.cov} has a similar form as that
of GR but with the values of the cosmological and gravitational
constants, $\Lambda$ and $\varrho^2$, specified by the boundary
conditions of the path integral (chosen to match their observed net
values). This result is in agreement with the canonical counting of
physical degrees of freedom: two propagating physical degrees of freedom
for the graviton and two zero modes. The zero modes are the
gravitational and cosmological constants.

Similar to the situation of unimodular gravity \cite{Bufalo:2015wda},
the local formulation of vacuum energy sequestering also involves
linearly dependent generators, namely, the constraints \eqref{cCi'} and
\eqref{cBi'} that are associated with the cosmological and
gravitational constants, respectively.  A proper treatment of
quantization and gauge invariance for a system with linearly dependent
generators is achieved by means of the Batalin-Vilkovisky
formalism \cite{Batalin:1984jr}. This was achieved in
Sect.~\ref{secCanVES.gf}, where we also showed that the path integral
obtained via the Batalin-Vilkovisky formalism matches that of the
reduced system obtained in Sect.~\ref{secCanVES.re} by eliminating
several unphysical variables.

Another similarity with unimodular gravity is the possibility of
generalizing the path integral expression in order to encompass
different vacuum states, which correspond to different values of the
gravitational and cosmological constants.
In the context of unimodular gravity, this approach leads to the
so-called Ng-van Dam form for the path integral, where an additional
integration over the cosmological constant is present.
We have extended the idea for local vacuum energy sequestering by
considering a superposition of vacuum states related to
different values of the gravitational and cosmological constants.
This results in a path integral with additional integration over both
the gravitational constant and the cosmological constant
\eqref{ZVESmatterJ}. The most likely values of the gravitational and
cosmological constants are regarded to contribute most to the path
integral. Then, in the semiclassical and stationary phase
approximation, the path integral implies a relation among the product
$\varrho^2\Lambda$ (or the ratio $\Lambda/G$) of the gravitational
constant and the cosmological constant to the average values of the
total pressure and the total energy density over the whole spacetime
\eqref{varrhoLambda}.

For completeness, we also considered the local vacuum energy
sequestering model from a new perspective of a topological or BRST-exact
formulation. In this approach, the gravitational action of vacuum energy
sequestering appears as a gauge-fixing action along with an appropriate
ghost action. As a result, the action of vacuum energy sequestering
becomes BRST exact and can be viewed as a topological field theory
[Eq.~\eqref{BRSTexact}]. The topological approach was supplemented by a
canonical analysis of the ghost sector, which is required in order to
establish the full BRST formalism.

Finally, we remark that the vacuum energy sequestering mechanism is
quite robust in its local form, and it could prove to be useful beyond
Einstein gravity. We noted the possibility of generalizing or modifying
the gravitational sector of the theory in a number of different ways.
For example, in order to achieve a power-counting renormalizable
gravitational sector, the Einstein-Hilbert part of the action could be
replaced with an action of the Ho\v{r}ava-Lifshitz type (with a
variable gravitational constant $\kappa^2$), which would be invariant
under foliation-preserving diffeomorphisms.
The possibility of modifying the terms that involve the auxiliary
fields, or the gauge conditions imposed on the auxiliary fields, was
also discussed within the BRST-exact formulation in Sect.~\ref{secIG}.
The latter could modify the sequestering mechanism drastically, and
hence it should be considered cautiously.

\subsection*{Acknowledgements}
M.O. gratefully acknowledges support from the Emil Aaltonen Foundation.
R.B. thankfully acknowledges CAPES/PNPD for support, Project No.
23038007041201166. The work of J.K. was supported by the Grant Agency of the Czech
Republic under the grant P201/12/G028.

\end{document}